\documentclass[prl,twocolumn,showpacs]{revtex4}
\usepackage{graphicx}
\begin{document}
%

%








%
\title{
Single Molecule Photon Statistics from a Sequence of Laser Pulses}
\author{F. Shikerman$^{1}$,  Y. He$^{1,2}$,  E. Barkai$^{1}$\\
$^{1}$Department of Physics, Bar Ilan University, Ramat-Gan 52900,
Israel\\$^2$School of Physics Science and Technology, Central South
University, Changsha 410083, China\\}
\begin{abstract}
{  There are many ways  of calculating photon statistics in quantum
optics in general and single molecule spectroscopy in particular
such as the generating function method, the quantum jump approach or
time ordering methods. In this paper starting with the optical Bloch
equation, within the paths interpretation of  Zoller, Marte and
Walls we obtain the photon statistics from a sequence of laser
pulses expressed by means of quantum trajectories. We find general
expressions for $P_n(t)$ - the probability of emitting n photons up
to time t, discuss several consequences and show that the
interpretation of the quantum trajectories (i) emphasizes
contribution to the photon statistics of the coherence paths
accumulated in the delay interval between the pulses and (ii) allows
simple classification of the terms negligible under certain physical
constraints . Applying this method to the concrete example of two
square laser pulses we find the probabilities of emitting 0,1 and 2
photons, examine several limiting cases and investigate the upper
and lower bounds of $P_0(t)$, $P_1(t)$ and $P_2(t)$ for a sequence
of two strong pulses in the limit of long measurement times.
Implication to single molecule non-linear spectroscopy and theory of
pairs of photons on demand are discussed briefly.}
\end{abstract}
\maketitle

\section{Introduction}

   The interaction of matter with a sequence of laser pulses is a powerful
tool frequently used for the investigation of a wide variety of
chemical, physical and biological systems \cite{MukamelB}. This
field of research called non-linear spectroscopy uses clever design
of laser pulses for the investigation of fast dynamics (e.g. pico -
seconds) of ensembles of molecules in the condensed phase. Recently
van Dijk et al \cite{vanDijk} reported the first experimental study
of an ultra-fast pump-probe single molecule system. Unlike the
previous approaches to such non-linear spectroscopy where only the
ensemble average response to the external fields is resolved, the
new approach yields direct information on single molecule dynamics,
gained through the analysis of photon counting statistics \cite{SB}.
In \cite{SB} we considered the non-linear spectroscopy for a single
molecule undergoing stochastic spectral diffusion process. Here we
neglect all dephasing and spectral diffusion effects, and
concentrate on the effect of the laser field parameters on the photon statistics.\\

   Another related application is the generation of two indistinguishable
photons using two short laser pulses interacting with a single
molecule or atom \cite{Santori}. Numerous applications for such
photon sources have been proposed
\cite{Hong,Knill,Shih,Katz,Bouwmeester} for the investigation of
entangled states between identical photons and quantum properties of
light, in the field of quantum information and quantum computation
requiring consecutive photons to have identical wave packets.
Usually in the mentioned experiments the single emitter is requested
to supply one or two photons within as short as possible time
interval. Although it is well established how to generate two
photons from two ideal $\pi$-pulses if the delay interval between
the pulses is very long, we can never produce two photons with
probability equal 1, when the interaction time is finite. Thus, the
information on the upper and lower bounds of the probabilities of
emitting 0, 1 and 2 photons as a function of the laser field
parameters, obtained
in the manuscript, can be very useful.\\

  Although the theory of single
particle photon statistics is well established \cite{Zoller,Mollow},
it remained unapplied due to the absence of experimental ability to
check the results. Recent experimental achievements
\cite{vanDijk,Santori,Katz,Orrit} allowed the investigation of the
interaction of a single quantum system with an external laser field
inspiring further development of theoretical methods
\cite{SB,BarkaiPRL,BarkaiRev,Mukamel,Cao,Xie,Goppich}. Today there
are several approaches to photon counting statistics such as
generating function method \cite{Brown} or quantum jump approach
\cite{Plenio} suitable for analytical predictions and numerical
calculations. In this paper we follow the path interpretation
approach of Mollow and Zoller, Marte and Walls \cite{Zoller,Mollow}
of the optical Bloch equations \cite{CT}, and show that this method
is very useful for the analysis
of single molecule non-linear spectroscopy.\\

 In what follows we consider a two level molecule interacting with two
laser pulses and obtain general expressions for $P_n(t)$ - the
probability of emitting n photons in interval $(0,t)$ by means of
quantum trajectories. We discuss the influence of the coherence on
the photon statistics. Also the explicit calculation of $P_0$, $P_1$
and $P_2$ - the probabilities of emitting 0,1 and 2 photons in the
limit of long measurement times $( t\to\infty )$ is investigated in
detail using the example of two identical square pulses. Some
technical details skipped in the text are given
in Appendixes A, B and C.\\

\section{Optical Bloch Equations}
   Interaction of an atom or a molecule with a radiation field is described
by the optical Bloch equations under well established conditions
\cite{CT}, and we remind the reader some of the basic assumptions.
First  (i)  the laser field is intense, so that it can be modeled
classically. Here the external electric
 field is ${\bf E}(t) = {\bf E_0} f(t)$, where
the amplitude ${\bf E_0}$ is independent of time.
(ii)
 The electronic states of the single emitter are modeled based on the two level
system approximation. This assumption is excellent when the laser is
resonating with a particular absorption frequency of the molecule,
the latter being well separated from other natural frequencies  of
the emitter. Most single molecules have a triplet state, however the
life time of the triplet is much longer than the time scales under
consideration in this manuscript, and it can be neglected. (iii) The
spontaneous emission process is described by the Markovian
approximation, where the inverse life time of the excited state is
$\Gamma$. (iv) We neglect thermal dephasing, spectral diffusion and
interaction of the emitter with a thermal bath, which was partially
treated in \cite{SB}. (v) Finally, we will assume that the dipole
moment of the excited and ground state of the single emitter is
zero, so that only the transition dipole moment of the particle is
important. Assumptions (i,ii,iii,iv) are physical assumptions which
are justified in many single molecule experiments at least at low
temperatures \cite{Orrit,Rozkov}, and condition (v) is not limiting
since our technique could be modified in principle to the case where
excited and ground
states of the molecule have a dipole.\\

   The two level system is described by a vector composed
of the density matrix elements: $\sigma=(\sigma_{\rm ee},
\sigma_{\rm gg}, \sigma_{\rm ge}, \sigma_{\rm eg})^{T}$. Here
$\sigma_{\rm ee}$ and $\sigma_{\rm gg}$ represent the populations of
the excited and ground states respectively and $\sigma_{\rm
ge},\sigma_{\rm eg}$ describe the coherences, namely the off
diagonal matrix elements of the density matrix, and obey
$\sigma_{\rm eg} ^{*}=\sigma_{\rm ge}$. The optical Bloch equation
is \cite{CT}
\begin{equation}
\dot{\sigma} = L\left( t \right) \sigma + \hat{\Gamma}\sigma,
\label{eqBloch}
\end{equation}
with
\begin{equation}
L(t) = \left(
\begin{array}{c c c c}
- \Gamma  & 0 & - i \Omega f(t)  & i \Omega f(t) \\
0 & 0 & i \Omega f(t) & - i \Omega f(t) \\
- i \Omega f(t) & i \Omega f(t) & i \omega_0 - \Gamma/2 & 0 \\
 i \Omega f(t) & - i \Omega f(t) & 0 & - i \omega_0 - \Gamma/2
\end{array}
\right) \label{eqOBE}
\end{equation}
and
\begin{equation}
\hat{\Gamma} = \left(
\begin{array}{c c c c}
0  & 0 & 0  & 0 \\
 \Gamma & 0 & 0  &  0 \\
 0 & 0  & 0  & 0 \\
 0 & 0  & 0  & 0
\end{array}
\right), \label{eqLoOBE}
\end{equation}\\
where the Rabi frequency is $ \Omega=-\frac{1 }{\hbar} {\bf d} _{ge}
\cdot {\bf E} _0 $ and ${\bf d} _{ge}$ is the transition dipole
moment of the two level system. The operator $\hat{\Gamma}$
Eq.~(\ref{eqLoOBE}) describes direct transition from the excited to
the ground state, and hence is associated with the spontaneous
emission of a single photon.\\

The optical Bloch equation Eq.~(\ref{eqBloch}) does not yield a
direct method for calculating the probabilities of the number of
emitted photons. However starting with \cite{Zoller,Mollow} an
interpretation of the optical Bloch formalism yields a tool for the
calculation of photon statistics, based on the n-photon-propagators
(see details below). The formal solution to Eq.~(\ref{eqBloch}) may
be given by the infinite iterative expansion in $\hat{\Gamma}$
\cite{Brown,Mukamel}:
$$ \sigma_{(t)} = {\cal G}(t,0) \sigma_{(0)} + \int_0 ^t {\rm d t_1} {\cal G} (t,t_1) \hat{\Gamma} {\cal G}(t_1, 0) \sigma_{(0)} + $$
\begin{equation}
+\int_0 ^{t} {\rm d} t_2 \int_0 ^{t_2} {\rm d} t_1 {\cal G}( t, t_2)
\hat{\Gamma} {\cal G}(t_2 , t_1) \hat{\Gamma} {\cal G} (t_1 , 0 )
\sigma_{(0)} + \cdots, \label{eqformal}
\end{equation}\\
where $\sigma_{(0)}$ is the initial condition, and the Green
function describing the evolution of the system in the absence of
spontaneous transitions into the ground state (i.e. without
$\hat{\Gamma}$ ) is
\begin{equation}
{\cal G} (t,t') = \hat{T} \exp\left[ \int_{t'} ^ t L(t_1) {\rm d}
t_1 \right],
\label{eqGreenf}
\end{equation}
where $\hat{T}$ is the time ordering operator. The first term in the
expansion Eq.~(\ref{eqformal}) does not include $\hat{\Gamma}$ at
all, and hence describes the process where no photons are emitted,
the second term includes $\hat{\Gamma}$ just once and describes the
process where one photon is emitted etc. It is therefore useful to
define the conditional state $\sigma^{(n)}_{(t)}$, where $n$ is an
index for the number of photons emitted in the time interval
$(0,t)$. Then by definition
\begin{equation}
 \sigma^{(n)}_{(t)} = U^{(n)}_{(t,0)} \sigma_{(0)},
\label{eqsign}
\end{equation}
where the n-photon-propagator \cite{Mukamel} is
\begin{equation}
 U^{(n)}_{(t,t')} =
\int_{t'} ^t {\rm d} t_n \cdots \int_{t'} ^{t_2} {\rm d} t_1~ {\cal
G}(t,t_n) \hat{\Gamma} \cdots\hat{\Gamma} {\cal G}(t_1,t') .
\label{eqUn}
\end{equation}
The physical origin of the n-photon-propagator defined by
Eq.~(\ref{eqUn}) is simple and intuitive: the system evolves
interacting with the laser field without photon emissions until time
$t_1$, it then emits a single photon and continues the evolution
without emissions until time $t_2$ when it emits the second photons
and so on.
   At this point it is convenient to choose a four-dimensional orthonormal basis to
work with: $| e \rangle=(1,0,0,0)^{T}$, $|g \rangle= (0,1,0,0)^{T}$,
$| c \rangle=(0,0,1,0)^{T}$ and $| c* \rangle= (0,0,0,1)^{T}$.
According to the matrix form of the Bloch equation
Eq.~(\ref{eqBloch}) the first two vectors correspond to pure excited
and ground states respectively. The last two vectors, however, do
not represent any real physical state and should be simply
considered as convenient mathematical way to include all possible
quantum paths going through superposition of the pure physical
states $|e\rangle$ and $|g\rangle$, thus representing the
contribution of the coherence effect. Using this notation the main
equation for calculating the probability of $n$ emission events up
to time $t$ is \cite{remark}
\begin{equation}
P_n(t) = (\langle e|+\langle g|)\sigma^{(n)}_{(t)} \rangle=(\langle
e|+\langle g|)U^{(n)}_{(t,0)}|\sigma_{(0)}\rangle. \label{eqPn}
\end{equation}
For example the probability of emitting zero photons is
\begin{equation}
P_0(t) = \sum_{i = e,g} \langle i | {\cal G} (t,0) |
\sigma_{(0)}\rangle,
\end{equation}
and the probability of emitting a single photon is
\begin{equation}
P_1(t) = \sum_{i = e,g} \langle i | \int_0 ^t {\rm d} t_1 {\cal G}
\left( t , t_1 \right) \hat{\Gamma} {\cal G} \left( t_1 , 0 \right)
| \sigma_{(0)} \rangle.
\end{equation}\\

   Consider a laser field interacting with the molecule in the time
interval $(t',t)$ and choose a fixed point $t_a$ inside this
interval. Such a partitioning of the time axis is useful for the
analysis of sequence of pulses investigated in the following
section, when we distinguish between time intervals where the laser
is turned on and off. First, let's split the integration over $t_n$
in Eq.~(\ref{eqUn}) into two parts:
\begin{equation}
 U^{(n)}_{(t,t')}=\int_{t'} ^{t}{\rm d} t_n\cdots=\int_{t'} ^{t_a}{\rm d}
t_n\cdots+\int_{t_a} ^{t}{\rm d} t_n\cdots.
\end{equation}
Using the fact that in the first interval $(t',t_a)$ $t_n \leq t_a$
and replacing the Green function ${\cal G}(t,t_n)$ by the product
${\cal G}(t,t_a){\cal G}(t_a,t_n)$ one easily finds
\begin{equation}
\int_{t'} ^{t_a}{\rm d} t_n\cdots \int_{t'} ^{t_2}{\rm d} t_1{\cal
G}(t,t_n) \hat{\Gamma}\cdots \hat{\Gamma} {\cal
G}(t_1,t')=U^{(0)}_{(t,t_a)}U^{(n)}_{(t_a,t')}.
\end{equation}
Now, left with the integral over the second range $(t_a,t)$, we
repeat exactly the same procedure as we did with the initial
expression, but this time we split the integration over $t_{n-1}$
into $(t',t_a)$ and $(t_a,t_n)$. Similarly, using $t_{n-1} \leq t_a$
and replacing ${\cal G}(t_n,t_{n-1})$ by ${\cal G}(t_n,t_a){\cal
G}(t_a,t_{n-1})$ inside the first interval we find
$$\int_{t_a} ^{t}{\rm d} t_n\int_{t'} ^{t_a}{\rm d} t_{n-1}\cdots
\int_{t'} ^{t_2}{\rm d} t_1{\cal G}(t,t_n) \hat{\Gamma}\cdots
\hat{\Gamma} {\cal G}(t_1,t')=$$
\begin{equation}
=U^{(1)}_{(t,t_a)}U^{(n-1)}_{(t_a,t')}.
\end{equation}
Repeating this algorithm n times it is easy to prove that
\begin{equation}
 U^{(n)}_{(t,t')}=\sum_{\alpha=0}^{n}U^{(\alpha)}_{(t,t_a)}U^{(n-\alpha)}_{(t_a,t')}.
\label{eqSumUU}
\end{equation}
Eq.~(\ref{eqSumUU}) means that the propagator corresponding to n
emission events in $(t',t)$ can be decomposed into $\alpha$ emission
events in $(t',t_a)$ and $n-\alpha$ emission events in $(t_a,t)$.
The extension to more than one time point such as $t_a$ is trivial
and leads to summation over all possible permutations of the n
photons propagators resulting in n emission events.\\

   Turning back to the Eq.~(\ref{eqPn}) for  $P_n(t)$ and inserting
 the closure relation
 \begin{equation}
\sum_{j=e,g,c,c*}|j \rangle\langle j|=1, \label{closure}
\end{equation}
 we find
\begin{equation}
P_n(t) = \sum_{i=e,g} \sum_{j=e,g,c,c*}  \sum_{\alpha=0} ^n \langle
i | U^{n-\alpha}_{(t,t_a)}| j \rangle \langle j|
U^{\alpha}_{(t_a,0)}| \sigma_{(0)} \rangle \label{eqPnt1}
\end{equation}
Eq.~(\ref{eqPnt1}) describes the summation over all possible paths
resulting in n emission events and suggests the following
classification : the paths going through the pure states
$|j\rangle=| e \rangle,| g \rangle$ may be identified as
semiclassical, whereas the paths going through the states
$|j\rangle=| c \rangle,| c^* \rangle$ describe the contribution of
the coherence.

\section{Two Pulses}

   Now we focus on the case of two laser pulses separated by a
window $\Delta$ in which the laser is turned off. The initial time
is $t=0$, the time $t_1$ is the moment when the first pulse is
switched off. The amplitude of the external field remains equal zero
$f(t)=0$ for the delay period $t_1 < t<t_1+\Delta$. At time $t_2=t_1
+ \Delta$ the laser is turned on again, and then again turned off
for $t>t_3$ ($f(t)=0$ for $t>t_3)$. Schematically the sequence is
represented in Fig.~1 for square pulses, however we emphasize that
the results obtained in this section are valid for pulses of any
shape. Our goal is the derivation of general expressions for
$P_n(t)$ from two pulses in the limit of the long measurement time
$t\rightarrow\infty$ when we know that eventually the system is in
the ground state. We assume that the molecule is always in the
ground state at the beginning of the experiment. If we divide the
time axis into four distinct intervals : two intervals when the
laser is turned on and two others when the laser is turned off, the
most general expression for $U^{(n)}_{(t,0)}$ following from the
extension of Eq.~(\ref{eqSumUU}) is
\begin{equation}
U^{(n)}_{(t,0)}=
U^{(n-\alpha-\beta-\gamma)}_{(t,t_3)}U^{(\gamma)}_{(t_3,t_2)}U^{(\beta)}_{(t_2,t_1)}U^{(\alpha)}_{(t_1,0)},
\label{eqPngen}
\end{equation}
where the superscripts $\alpha$, $\beta$ and $\gamma$ are all non
negative integer values leading to n photons (i.e.~
$n-\alpha-\beta-\gamma\geq0$). The Einstein's summation rule from 0
to n must be applied to every superscript appearing twice. Inside
time intervals $(t,t_3)$ and $(t_2,t_1)$, when the laser is turned
off, the Rabi frequency is equal zero $\Omega=0$, and the
calculation of the Green function ${\cal G} (t,t')$
Eq.~(\ref{eqGreenf}) becomes nearly trivial. For the delay interval
$\Delta$ we find only two non-zero n-photon-propagators:
\begin{equation}
 U^{(0)}_{(t_1+\Delta,t_1)} =\left(
\begin{array}{c c c c}
e^{-\Gamma\Delta}  & 0 & 0  & 0 \\
 0 & 1 & 0  &  0 \\
 0 & 0  & e^{ (i \omega_0 - \Gamma/2)\Delta}  & 0 \\
 0 & 0  & 0  &   e^{ -(i \omega_0 + \Gamma/2)\Delta}
\end{array}
\right),\label{eqU0}
\end{equation}
\begin{equation}
 U^{(1)}_{(t_1+\Delta,t_1)} =\left(
\begin{array}{c c c c}
 0 & 0 & 0  & 0 \\
 1-e^{-\Gamma\Delta} & 0 & 0  &  0 \\
 0 & 0  & 0  & 0 \\
 0 & 0  & 0  & 0
\end{array}
\right) \label{eqU1}\end{equation}
and $U^{(n)}_{(t_1+\Delta,t_1)}=0$ for $n>1$. This result is
definitely expected , since if nothing excites the molecule, there
is no chance to get more than a single photon. The matrix
representation of the propagators in the interval $(t>t_3)$ when
$t\rightarrow\infty$, are found by taking the limit
$\Delta\rightarrow\infty$ of Eqs.~(\ref{eqU0}),(\ref{eqU1}). Now
inserting the closure relation Eq.~(\ref{closure}) between each two
propagators of Eq.~(\ref{eqPngen}) and using the just obtained
matrix elements Eqs.~(\ref{eqU0}),(\ref{eqU1}) we find :
\begin{widetext}
\begin{equation} P_n=\lim_{t\rightarrow\infty}P_n(t)=P_n^{\rm Cla}+
e^{-\Delta\frac{\Gamma}{2}}(e^{i \Delta\omega_0}A_n^{\rm
Coh}+C.C.),\label{eqPn1}\end{equation}
where
$$P_n^{\rm Cla}=\sum_{\alpha=0}^{n}\left\{\langle g|U^{(n-\alpha)}_{(t_3,t_2)}|g\rangle
\langle g|U^{(\alpha)}_{(t_1,0)}|g\rangle  +
e^{-\Delta\Gamma}\langle g|U^{(n-\alpha)}_{(t_3,t_2)}|e\rangle
\langle
e|U^{(\alpha)}_{(t_1,0)}|g\rangle\right\}+\sum_{\alpha=0}^{n-1}(1-e^{-\Delta\Gamma})\langle
g|U^{(n-\alpha-1)}_{(t_3,t_2)}|g\rangle \langle
e|U^{(\alpha)}_{(t_1,0)}|g\rangle$$
\begin{equation}
 + \sum_{\alpha=0}^{n-1}\left\{\langle
e|U^{(n-\alpha-1)}_{(t_3,t_2)}|g\rangle \langle
g|U^{(\alpha)}_{(t_1,0)}|g\rangle +e^{-\Delta\Gamma}\langle
e|U^{(n-\alpha-1)}_{(t_3,t_2)}|e\rangle \langle
e|U^{(\alpha)}_{(t_1,0)}|g\rangle\right\}+\sum_{\alpha=0}^{n-2}(1-e^{-\Delta\Gamma})\langle
e|U^{(n-\alpha-2)}_{(t_3,t_2)}|g\rangle \langle
e|U^{(\alpha)}_{(t_1,0)}|g\rangle\label{Pncla}\end{equation}
and
\begin{equation}A_n^{\rm Coh}=\sum_{\alpha=0}^{n}\langle
g|U^{(n-\alpha)}_{(t_3,t_2)}|c\rangle \langle
c|U^{(\alpha)}_{(t_1,0)}|g\rangle  + \sum_{\alpha=0}^{n-1}\langle
e|U^{(n-\alpha-1)}_{(t_3,t_2)}|c\rangle \langle
c|U^{(\alpha)}_{(t_1,0)}|g\rangle.\label{Ancoh}\end{equation}
\end{widetext}

 Eqs.~(\ref{eqPn1}),(\ref{Pncla}) and (\ref{Ancoh}) summarize all possible
paths resulting in n photon emission events and allow simple
identification of negligible terms, when particular physical
constraints are taken into account. It is easy to see, that the
first two terms of $P_n^{\rm Cla}$ Eq.~(\ref{Pncla}) (those with
$\sum_{\alpha=0}^{n}$) describe processes where all n photons are
emitted during the pulses and none in the delay interval or after
the second pulse. The third term of this expression represents the
processes, where a single photon is emitted in the delay period
(with probability $\left[1-e^{-\Delta\Gamma}\right]$) and n-1
photons during the pulses events. Similarly, the next two terms
originate from the processes, where a single photon is emitted after
the second pulse and zero photons in the delay interval, and
finally, the last term describes situations, where one photon is
emitted in the delay interval and another after the second pulse.
This interpretation may be used to simplify the calculations, as for
instance in the case of the short pulses considered below, where we
neglect the trajectories with photons emitted during the pulse events.\\

    Although Eqs.~(\ref{eqPn1}),(\ref{Pncla}) and (\ref{Ancoh}) are very
general, they already contain interesting physical information.
First of all, we pay attention to the fact, that the coherence terms
$A_n^{\rm Coh}$, describing the processes where the molecule is left
in the superposition of the pure states at the end of the first
pulse (i.e. the paths going trough the states $|c\rangle$ and
$|c^*\rangle$), never include trajectories where a photon is emitted
within the delay interval $\Delta$. Mathematically this follows from
Eqs.~(\ref{eqU0}) and (\ref{eqU1}), and physically it makes sense,
because spontaneous collapse into the ground state destroys the
coherence. Secondly, we see, that the coherence terms are multiplied
by the exponentially decaying factor $e^{-\Gamma\Delta/2}$,
responsible for the dephasing effect, and oscillate in $\Delta$ with
orbital frequency $\omega_0$ (see the $e^{i\Delta\omega_0}$ term in
Eq.~(\ref{eqPn1})). In optics $\omega_0$ is much larger than the
inverse of $\tau$ - the minimum time resolution of the measurement
device : $\tau\omega_0\gg1$. Therefore, in order to match our
results for the probability of emitting n photons to those observed
by an experimentalist, it is essential to treat the coherence terms
as stochastic variables - i.e. it is reasonable to replace them with
their time average, which is equal zero. However, it should not be
forgotten that : (i) in the limit $\Delta\to0$, when the pulses are
attached together, the coherence contribution $A_n^{\rm Coh}$
becomes non-oscillating and non-negligible part of $P_n$, and (ii)
in non-optical microwave experiments, where the absorption frequency
is comparable with the time resolution of the measurement device
\cite{Katz},
the influence of the coherence trajectories is important.\\

   It is possible to derive another useful expression for the probability
of emitting n photons from two pulses. First we note, that according
to Eqs.~(\ref{eqPn}), (\ref{eqU0}) and (\ref{eqU1}) the probability
of emitting n photons from any single pulse or sequence of pulses of
total length T in the limit of infinitely long measurement time
$t\to\infty$ is
\begin{equation}
P_n=\langle g|U^{(n)}_{(T,0)}|g\rangle + \langle
e|U^{(n-1)}_{(T,0)}|g\rangle, \label{SinglePulse}
\end{equation}
(where for n=0 the second term $\langle
e|U^{(-1)}_{(T,0)}|g\rangle=0$). From the physical point of view the
second term of Eq.~(\ref{SinglePulse}) expresses the fact, that the
molecule left in the pure excited state eventually decays to the
ground state by spontaneous photon emission. Simple rearrangement of
Eqs.~(\ref{Pncla}) and (\ref{Ancoh}), with details given in Appendix
A, results in :
\begin{widetext}
\begin{equation}
P_n=\sum_{\alpha=0}^{n}P_{n-\alpha}^{I_2}P_{\alpha}^{I_1}+
e^{-\Delta\Gamma}\left\{P_{n}^{I_2I_1}-\sum_{\alpha=0}^{n}P_{n-\alpha}^{I_2}P_{\alpha}^{I_1}
+\left[(e^{\Delta(\Gamma/2+i\omega_0)}-1)A_n^{\rm
Coh}+C.C.\right]\right\}. \label{Pncorrel}
\end{equation}
\end{widetext}
Here $P_n^{I_1}=\langle g|U^{(n)}_{(t_1,0)}|g\rangle + \langle
e|U^{(n-1)}_{(t_1,0)}|g\rangle$ is the probability of emitting n
photons only from the first pulse \cite{Yong,Yong1}, similarly
$P_n^{I_2}=\langle g|U^{(n)}_{(t_3,t_2)}|g\rangle + \langle
e|U^{(n-1)}_{(t_3,t_2)}|g\rangle$ designates the probability of
emitting n photons only from the second pulse, and
\begin{equation}
P_n^{I_2I_1}=\sum_{\alpha=0}^n\langle
g|U^{(n-\alpha)}_{(t_3,t_2)}U^{(\alpha)}_{(t_1,0)}|g\rangle +
\sum_{\alpha=0}^{n-1}\langle
e|U^{(n-1-\alpha)}_{(t_3,t_2)}U^{(\alpha)}_{(t_1,0)}|g\rangle
\label{Pi2i1} \end{equation}
 is the probability of emitting n
photons from the two pulses produced one immediately after another
(i.e. with zero delay). This formulation of $P_n$
Eq.~(\ref{Pncorrel}) shows, that the first term
$\sum_{\alpha=0}^{n}P_{n-\alpha}^{I_2}P_{\alpha}^{I_1}$ represents
the sum of all possible ways of emitting n photons from the both
pulses, as if the consequences of the interaction of the molecule
with the first pulse had no influence on the state of the system at
the beginning of the second pulse, i.e. like if the treatment of
each pulse could be done independently. Nevertheless, since such an
influence exists, it is reasonable to define the rest of the terms
on the righthand side of Eq.~(\ref{Pncorrel}) as a correlation:
$$C(\Delta)=P_n-\sum_{\alpha=0}^{n}P_{n-\alpha}^{I_2}P_{\alpha}^{I_1}=
e^{-\Delta\Gamma}\left\{P_{n}^{I_2I_1}-\sum_{\alpha=0}^{n}P_{n-\alpha}^{I_2}P_{\alpha}^{I_1}+\right.$$
\begin{equation}
\left.+\left[(e^{\Delta(\Gamma/2+i\omega_0)}-1)A_n^{\rm
Coh}+C.C.\right]\right\}. \label{correl} \end{equation}
Note that Eq.~(\ref{Pncorrel}) makes perfect physical sense in the
limits $\Delta\rightarrow\infty$ and $\Delta\rightarrow0$ where we
find trivially expected results. In the first case only the first
term on the righthand side of Eq.~(\ref{Pncorrel}) survives - i.e.
this limit describes the situation where the interaction of the
molecule with the first pulse indeed has no influence on the
interaction of the molecule with the second pulse, since all
coherence effects have enough time to decay completely. And the
second limit gives $P_n=P_{n}^{I_2I_1}$. We emphasize, that in the
first case, once the probabilities of emitting n photons from each
single pulse are known, the efforts needed for the calculations are
considerably reduced. However, care must be taken while using
Eq.~(\ref{Pncorrel}) for the calculation of the second limit
$\Delta\rightarrow0$, since the continuity of the laser's phase
plays important role, as demonstrated on the example of two square
pulses in the subsequent section.

\section{Example : Two Square Pulses}
   In this section we apply our general results to the concrete example of two identical square laser pulses.
Consider the sequence :
\begin{figure}
\includegraphics[width=\columnwidth]{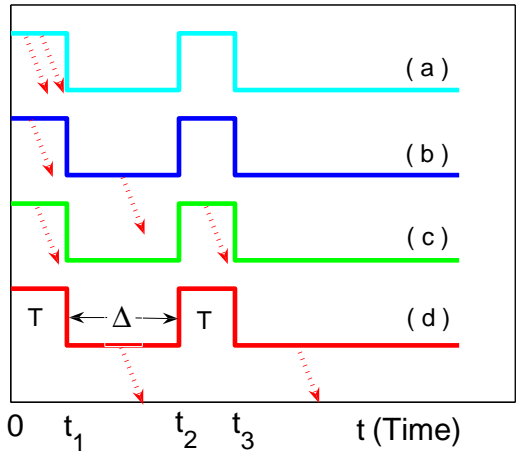}\\
  \caption{Two square laser pulses as modeled in Eq.~(\ref{Figure1}).
T is the length of a single pulse and $\Delta$ is the delay interval
between the pulses. The dashed arrows show four cases out of eight
possibilities of emission of two photons. For example: (a) two
photons are emitted within the duration of the first
pulse.}\label{model1}
\end{figure}
\begin{equation}
f(t) = \left\{
\begin{array}{l l}
\cos(\omega_L t + \phi_1 ) & \ \ 0<t<t_1\\
0 & \ \ t_1< t<t_2\\
\cos\left[ \omega_L (t-t_2) + \phi_2  \right]  & \ \ t_2< t<t_3 \\
0 & \ \ t_3 < t
\end{array}
\right.
 \label{Figure1}
\end{equation}
where $t_1=t_3-t_2=T$ - is the pulse's duration, $t_2-t_1=\Delta$ -
is the delay period between the pulses and $\phi_1$ and $\phi_2$ are
the initial phases of each pulse. For simplicity we assume, that the
laser frequency $\omega_L = \omega_0$, namely we consider the case
of zero detuning, and also the initial phase of the first pulse is
zero $\phi_1=0$. The time dependence of the Rabi frequency $\Omega$,
describing the interaction, is schematically illustrated in Fig.~1.
The propagators acting on the molecule during the square pulses are
found using the Rotating Wave Approximation (RWA). In Appendixes B
and C we show, that within RWA Eq.~(\ref{eqPn1}) can be rewritten in
the form
\begin{equation}P_n=p_n^{\rm Cla}+
e^{-\Delta\frac{\Gamma}{2}}(e^{i (\Delta+T)\omega_{0}-\phi_{2}}a_n^{\rm Coh}+C.C.),\label{eqPnRWA}\end{equation}\\
where the definition of $p_n^{\rm Cla}$ and $a_n^{\rm Coh}$ follows
from $P_n^{\rm Cla}$ and $A_n^{\rm Coh}$
Eqs.~(\ref{Pncla}),(\ref{Ancoh}) by replacing  all
n-photon-propagators Eq.~(\ref{eqUn}) by $\tilde U^{(n)}_{(t,t')}$ -
the n-photon-propagators calculated within RWA (see the derivation
of Eq.~(\ref{eqtildUn}) in Appendix B) . All mathematical
manipulations and calculations
were obtained with the help of Mathematica 5.0.  \\

   First we consider in detail the probability of emitting zero photons.
For the calculation of $P_0(t)$ we need the zero photon propagator.
This is the simplest case, since there is only one possible
permutation. According to Eq.~(\ref{eqPngen}) we have :
\begin{equation}
U^{(0)}_{(t,0)}=U^{(0)}_{(t,t_3)}U^{(0)}_{(t_3,t_2)}U^{(0)}_{(t_2,t_1)}U^{(0)}_{(t_1,0)}.
\label{eqP0}
\end{equation}
Inserting the closure relation Eq.~(\ref{closure}) and applying RWA
lead to :
$$p^{\rm Cla}_0 = e^{ -
\Gamma \Delta}\langle {\rm g}|\tilde U^{(0)}_{(t_3,t_2)}| {\rm e}
\rangle \langle {\rm e} |\tilde  U^{(0)}_{(t_1,0)} | {\rm
g}\rangle+$$
\begin{equation}
 + \langle {\rm g}|\tilde U^{(0)}_{(t_3,t_2)}| {\rm g} \rangle
\langle {\rm g}|\tilde U^{(0)}_{(t_1,0)}| {\rm g} \rangle
,\label{pzeroCla}
\end{equation}

\begin{equation}
a^{\rm Coh}_0=\langle {\rm g} | \tilde  U^{(0)}_{(t_3,t_2)} | {\rm
c} \rangle \langle c |\tilde U^{(0)}_{(t_1,0)}| {\rm g} \rangle,
\label{pzeroCoh}
\end{equation}\\
Calculating the matrix elements of Eqs.~(\ref{pzeroCla}) and
(\ref{pzeroCoh}) we find the following explicit expression for
$P_0=\lim_{t\to\infty}P_0(t)$:
\begin{widetext}
$$P_0 = {16 \Omega^4 e^{  - T - \Delta} \over \left( 1 - 4 \Omega^2\right)^2} \sinh^4 \left( { \sqrt{1 - 4 \Omega^2} \over 4} T \right) + $$
$$  {e^{-T} \over \left( 1 - 4 \Omega^2\right)^2} \left[ \left( 1 - 2 \Omega^2\right) \cosh\left( { \sqrt{ 1 - 4 \Omega^2}  \over 2} T \right) +
\sqrt{1 - 4 \Omega^2} \sinh\left( {\sqrt{1 - 4 \Omega^2} \over 2} T
\right) - 2 \Omega^2 \right]^2 $$
\begin{equation}
- {8 \Omega^2 \over \left(1 - 4 \Omega^2\right)^2 } e^{ - T -
\Delta/2 } \sinh^2\left({\sqrt{1 - 4 \Omega^2} \over 4} T \right)
\left[ \sqrt{ 1 - 4 \Omega^2} \cosh\left( { \sqrt{ 1 - 4 \Omega^2}
\over 4 } T \right) + \sinh\left( { \sqrt{1 - 4 \Omega^2} \over 4 }
T \right) \right]^2 \cos[\omega_0
\left(\Delta+T\right)-\phi_2],\label{P0exact}
\end{equation}
where we set $\Gamma=1$ for simplicity. The last term, exhibiting
oscillations due to the $\cos[\omega_0
\left(\Delta+T\right)-\phi_2]$, results from the quantum paths going
through the $|c\rangle$ and $|c*\rangle$, thus representing the
coherence effect. Of course, when $T=0$ or $\Omega=0$, $P_0=1$ since
no photons are emitted, and if $T \to \infty$, $P_0 =0$ since many
photons are emitted. Similar calculations were made also for $P_1$
and $P_2$ - see Eqs.~(\ref{P1}), (\ref{P2}) for the
final results in Appendix C.\\

    For very intense laser fields, when the Rabi frequency is much larger than the inverse life time
of the excited state, taking the limit $\Omega \gg1$ of
Eq.~(\ref{P0exact}) we obtain :
\begin{equation}
\lim_{\Omega\gg1}P_0  \sim e^{ - T} \left\{  e^{-\Delta}
\sin^4\left( { \Omega T \over 2} \right)  + \cos^4 \left( { \Omega T
\over 2} \right) - {1 \over 2} e^{ - \Delta/2} \sin^2 \left( \Omega
T \right) \cos [\omega_0 \left(\Delta+T\right)-\phi_2]\right\}.
\label{eqppzz}
\end{equation}
And using Eqs.~(\ref{P1}), (\ref{P2}):
$$\lim_{\Omega \gg1}P_1 = \frac{e^{-T}}{8}  \left \{4 \sin^2(\Omega T)+2 T \left[1-\sin^2\left(\frac{\Omega T}{2}\right)\left(1-e^{-\Delta}\right) \right]+T \sin^2(\Omega T)\left(1+e^{-\Delta}\right) \right\} $$
\begin{equation}
+  \frac{e^{-(T+{\Delta/2})}}{4} (T+2)  \sin ^2(\Omega T)
\cos[\omega_0 \left(\Delta+T\right)-\phi_2], \label{P1exact}
\end{equation}
$$\lim_{\Omega \gg1} P_2  = e^{-T} \left \{\left[ \sin ^4\left(\frac{\Omega T}{2}\right)+\frac{ T^2}{64}\cos ( \Omega T)  \right] \left(1-e^{-\Delta}\right) + \frac{T}{4} \left[\cos ^2(\Omega T)+2\right]\right \}$$
\begin{equation}
+\frac{ e^{-T} }{32} T^2 \left( \cos ^2(\Omega T)+4
\right)\left(1+e^{-\Delta}\right)-\frac{e^{-(T+{\Delta/2})}}{16}  T
(T+4) \sin ^2(\Omega T)  \cos [\omega_0
\left(\Delta+T\right)-\phi_2]. \label{P2exact}
\end{equation}\\
\end{widetext}

  Finally, we would like to  investigate the limiting behavior of
$P_0$, $P_1$ and $P_2$ within the strong fields approximation in the
case of long $\Delta\rightarrow\infty$ and short
$\Delta\rightarrow0$ delay intervals. As shown in \cite{Yong,Yong1},
the probabilities of emission 0, 1 and 2 photons from a single
square pulse of length T (see Eq.~(\ref{SinglePulse})) are given by
\begin{equation}
\lim_{\Omega\gg1}P_0^{I}=\lim_{\Omega\gg1}P_0^{I_1}=\lim_{\Omega\gg1}P_0^{I_2}\sim
e^{ - T/2} \cos^2 \left( { \Omega T \over 2} \right) ,
\label{p0single}
\end{equation}
\begin{equation}
\lim_{\Omega\gg1}P_1^{I}\sim \frac{e^{ - T/2}}{8}\left[
4+2T-(4+T)\cos(\Omega T ) \right]  \label{p1single}
\end{equation}
and
\begin{equation}
\lim_{\Omega\gg1}P_2^{I}\sim \frac{e^{ - T/2}}{64}T\left[
4T+16+(8+T)\cos(\Omega T ) \right] . \label{p2single}
\end{equation}
Taking the limit $\Delta\gg1$ of Eq.~(\ref{P0exact}) we find
\begin{equation}
\lim_{\Omega\gg1,\Delta\gg1}P_0=(P_0^{I})^2\sim e^{ - T} \cos^4
\left( { \Omega T \over 2} \right) , \label{p0deltainfty}
\end{equation}
which is equal precisely to the product of the probabilities of
emitting zero photons from two single square pulses
Eq.~(\ref{p0single}) and completely agrees with
Eq.~(\ref{Pncorrel}). Now using Eqs.~(\ref{Pncorrel}),
(\ref{p1single}), (\ref{p2single}) and the fact, that the two pulses
are identical, we can easily obtain the limit
$\Delta\rightarrow\infty$ of $P_1$ and $P_2$:
$$\lim_{\Omega\gg1,\Delta\gg1}P_1=2P_1^IP_0^I=$$
\begin{equation}
\frac{e^{ - T}}{4} \left[ (8+3T)\cos^2 \left( { \Omega T \over 2}
\right)-2(4+T)\cos^4 \left( { \Omega T \over 2} \right)\right]
\label{p1deltaifnty}
\end{equation}
and
$$\lim_{\Omega\gg1,\Delta\gg1}P_2=2P_0^{I}P_2^{I}+(P_1^{I})^2=\frac{e^{-T}}{64}\left[24 + T(40 + 9T) +\right.$$
\begin{equation}\left.(-32 +
T^2)\cos(\Omega T) +(8 + T(8 + T))\cos(2\Omega T)\right].
\label{p2deltaifnty}
\end{equation}\\

  Considering the opposite limit $\Delta\to0$ we remind, that the contribution of the coherence
paths going through the states $|c\rangle$ and $|c^*\rangle$ must
not be neglected. Moreover, it is essential to take into account,
that two attached square pulses of the same Rabi frequency are equal
to a single long square pulse, only if
$\phi_2=\phi_1+\omega_{0}(\Delta+T)$ - i.e. the pulse is continuous.
Assuming for simplicity, that this is the case, when
$\Delta\rightarrow0$ from Eq.~(\ref{P0exact}) we find for $P_0$ :
\begin{equation}
\lim_{\Omega\gg1,\Delta\ll1}P_0\sim e^{ - T}  \cos^2(\Omega T) ,
\label{p0deltazero}
\end{equation}
which once again agrees with Eq.~(\ref{Pncorrel}), since it is
exactly the result of replacing T with 2T in Eq.~(\ref{p0single}).
Similarly the limits $\Delta\to0$ of $P_1$ and $P_2$ may be obtained
by replacing T with
2T in Eqs.~(\ref{p1single}) and (\ref{p2single}).\\

 In Fig.~2, neglecting the fast oscillating coherence paths, we
plotted the semiclassical terms  $p_0^{\rm Cla}$,  $p_1^{\rm Cla}$
and $p_2^{\rm Cla}$ for the relatively long $\Delta=3$ and short
$\Delta=0.5$ delay intervals for the case of strong laser field
$\Omega=10$. 
Comparing the graphs one may see, that the dependence of $p_1^{\rm
Cla}$ on $\Delta$ is visibly weaker than those of $p_0^{\rm Cla}$
and $p_2^{\rm Cla}$ (we explain this effect later - see the
discussion below Table 1). When $\Gamma T\gg1$, we expect that : (i)
$p_0^{\rm Cla}, p_1^{\rm Cla}$ and $p_2^{\rm Cla}$ are all small,
since many photons are expected to be emitted during the pulses, and
(ii) independent of $\Delta$, since the contribution of the photons
emitted during the pulse event is much larger than the contribution
of the photons emitted in the delay interval. Such a behavior is
clearly seen for $p_2^{\rm Cla}$ (compare Figs.~$2.{\bf c}$ and
$2.{\bf f}$) where the difference between the case $\Delta=3$ and
$\Delta=0.5$ is stronger for short T. Below we prove this in a
Poissonian limit.

\begin{widetext}

\begin{figure}
\begin{eqnarray*}
 \includegraphics[height=1.5in, width=2in]{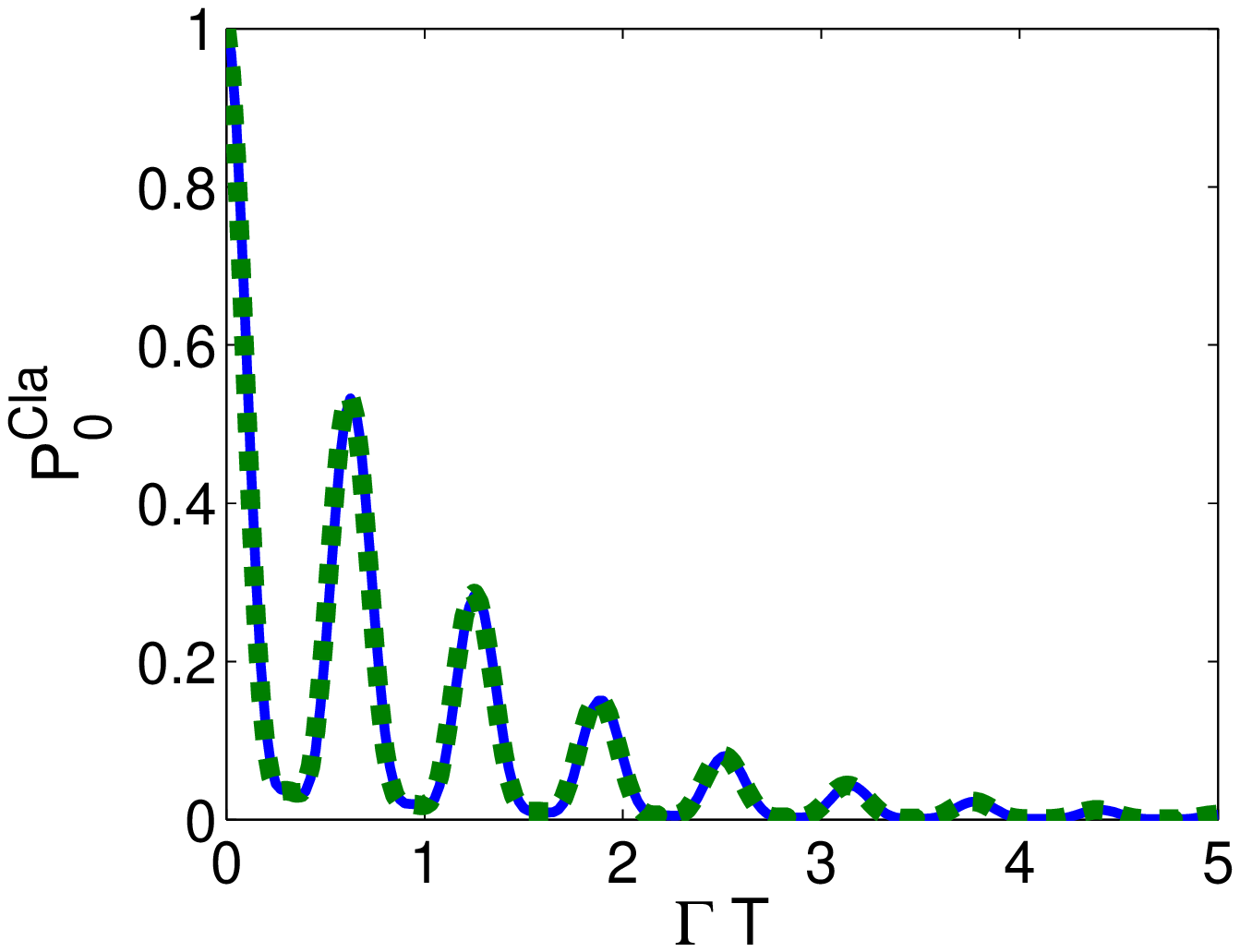}\textbf{a}&
 \includegraphics[height=1.5in, width=2in]{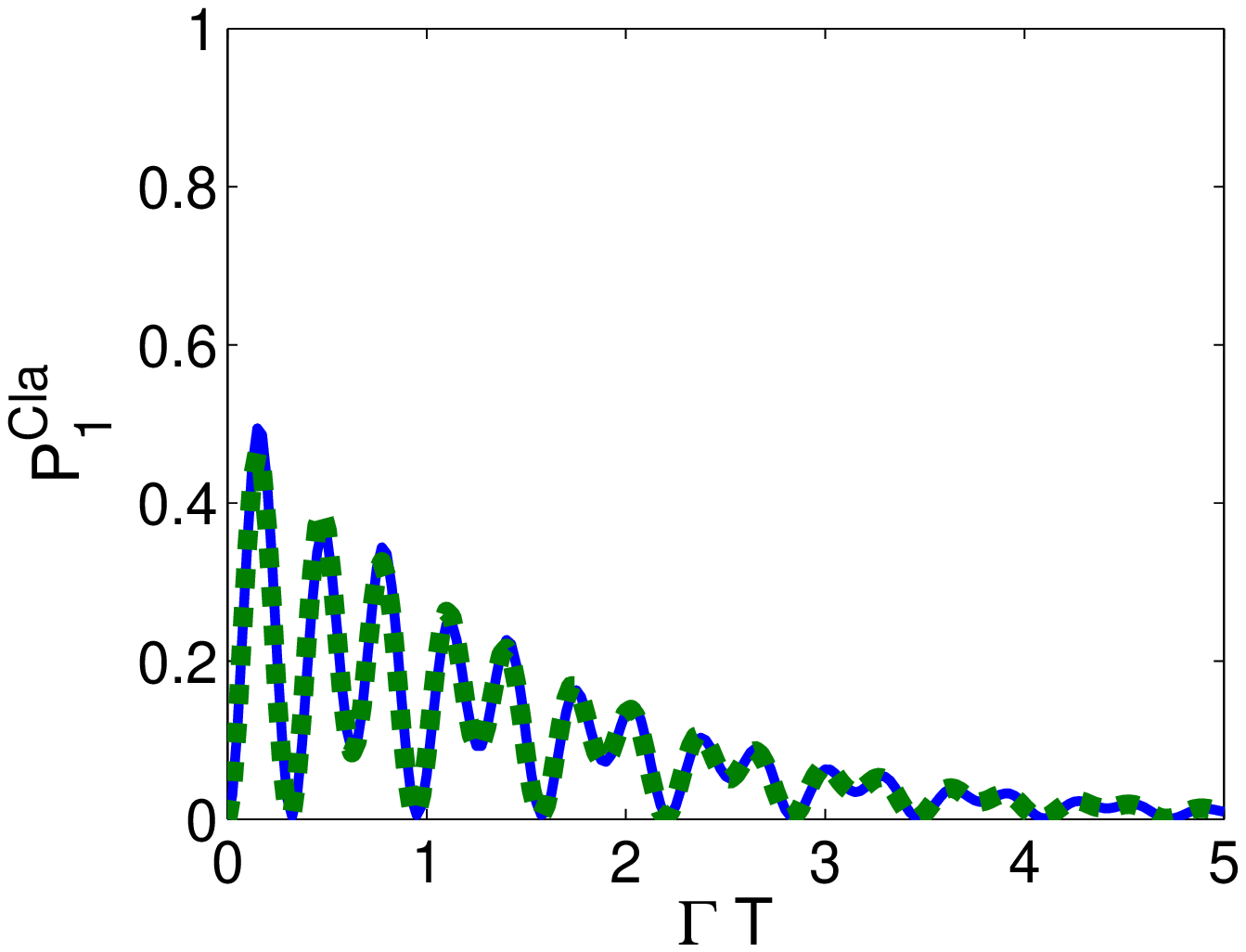}\textbf{b}&
  \includegraphics[height=1.5in, width=2in]{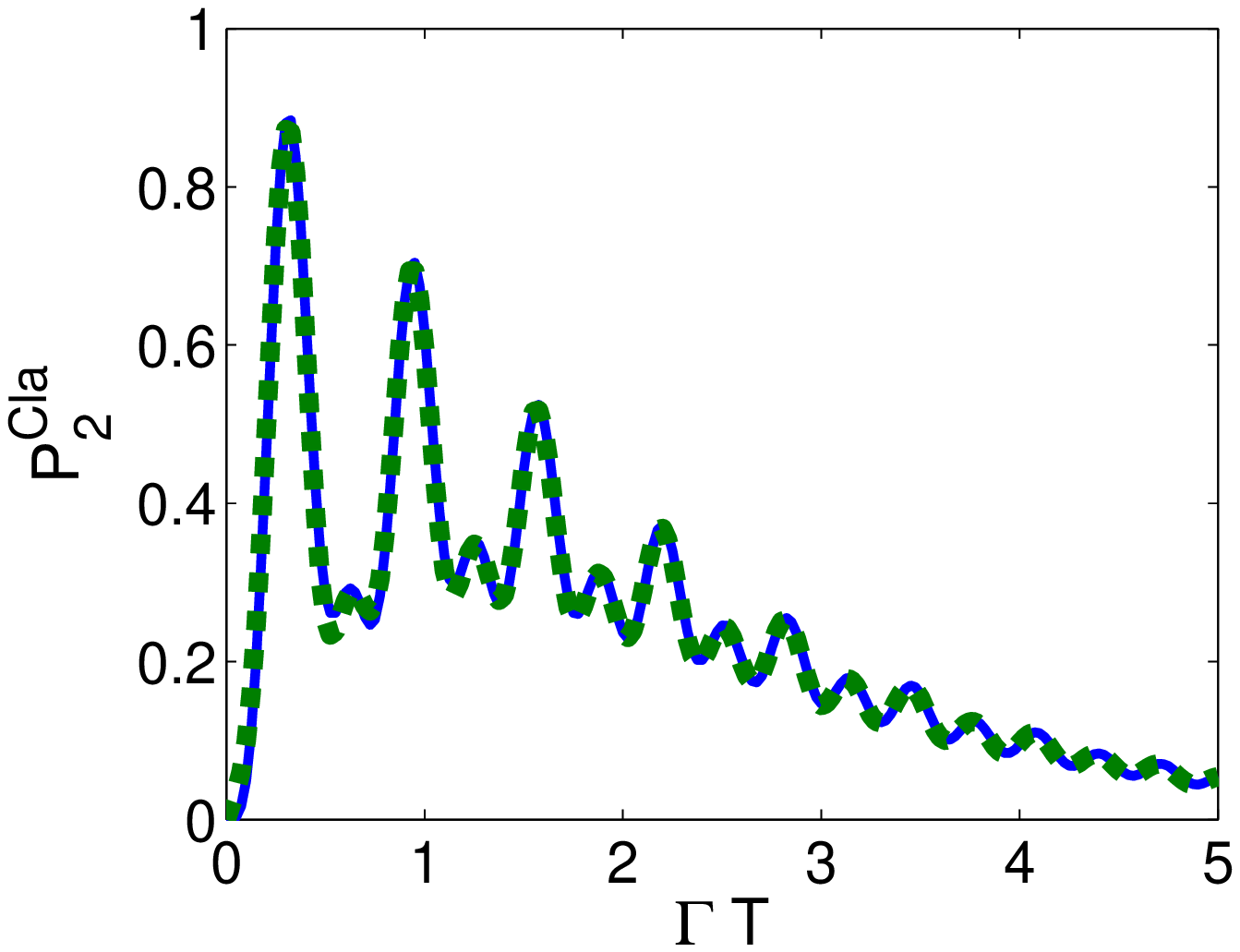}\textbf{c}\\
  \includegraphics[height=1.5in, width=2in]{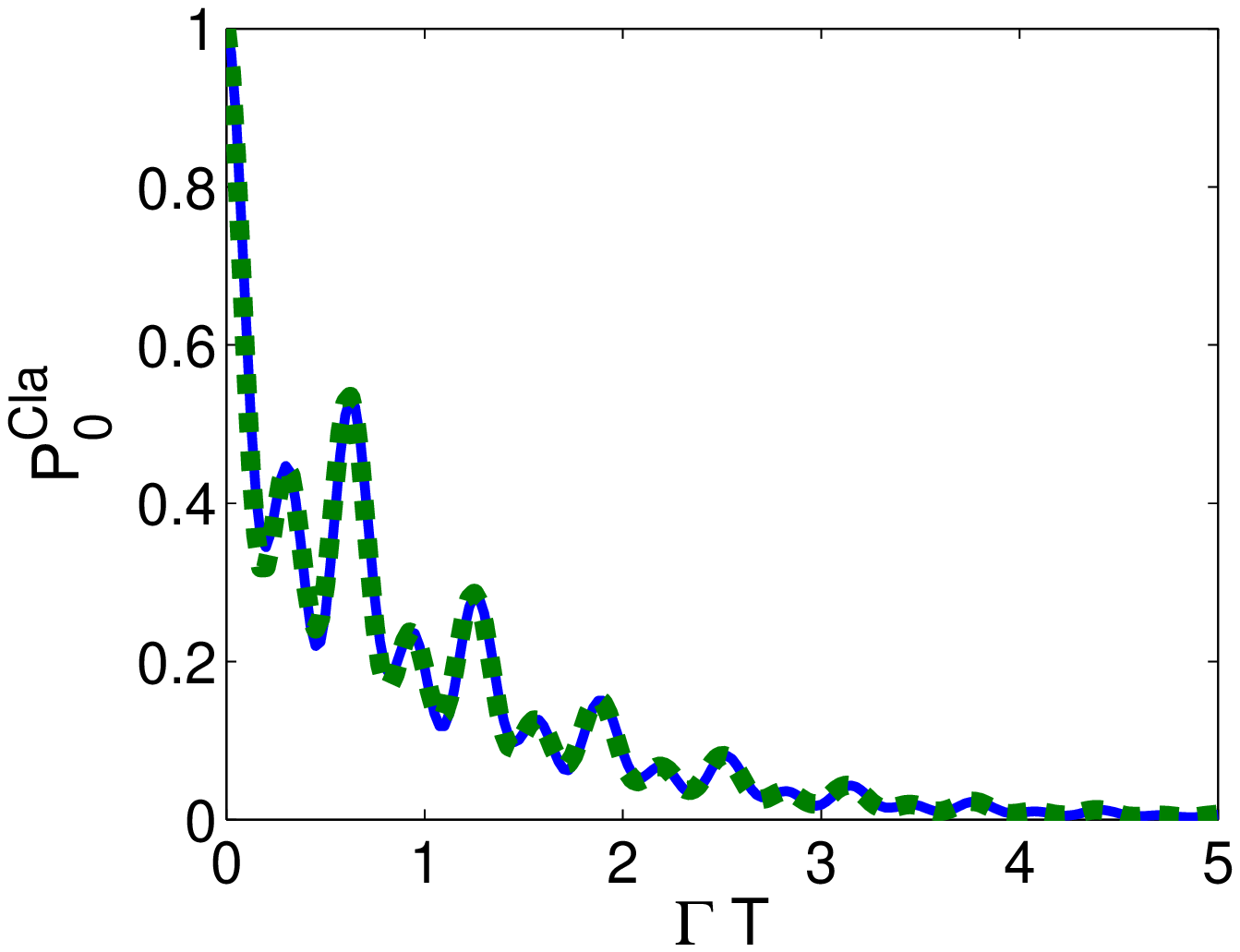}\textbf{d}&
 \includegraphics[height=1.5in, width=2in]{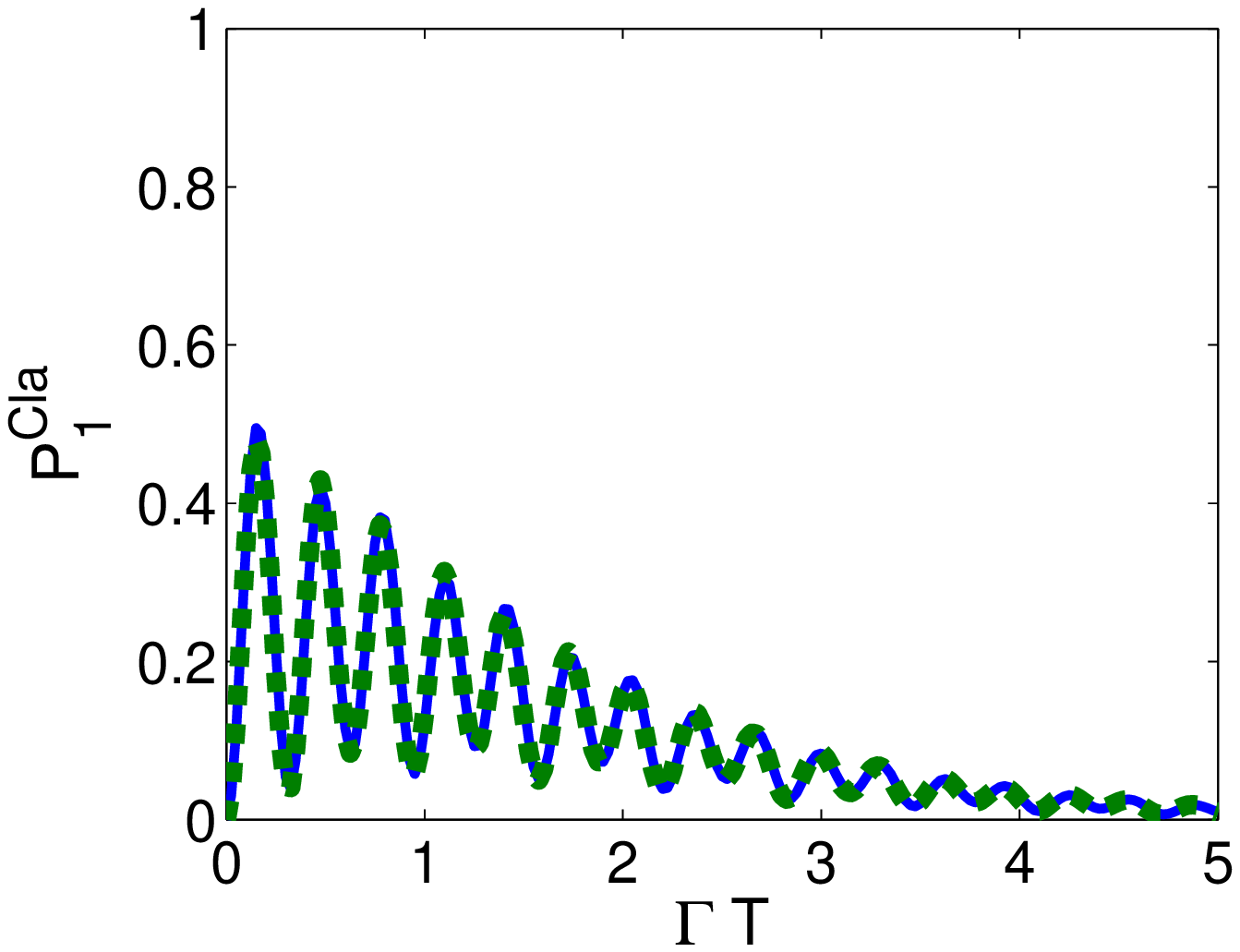}\textbf{e}
 &
  \includegraphics[height=1.5in, width=2in]{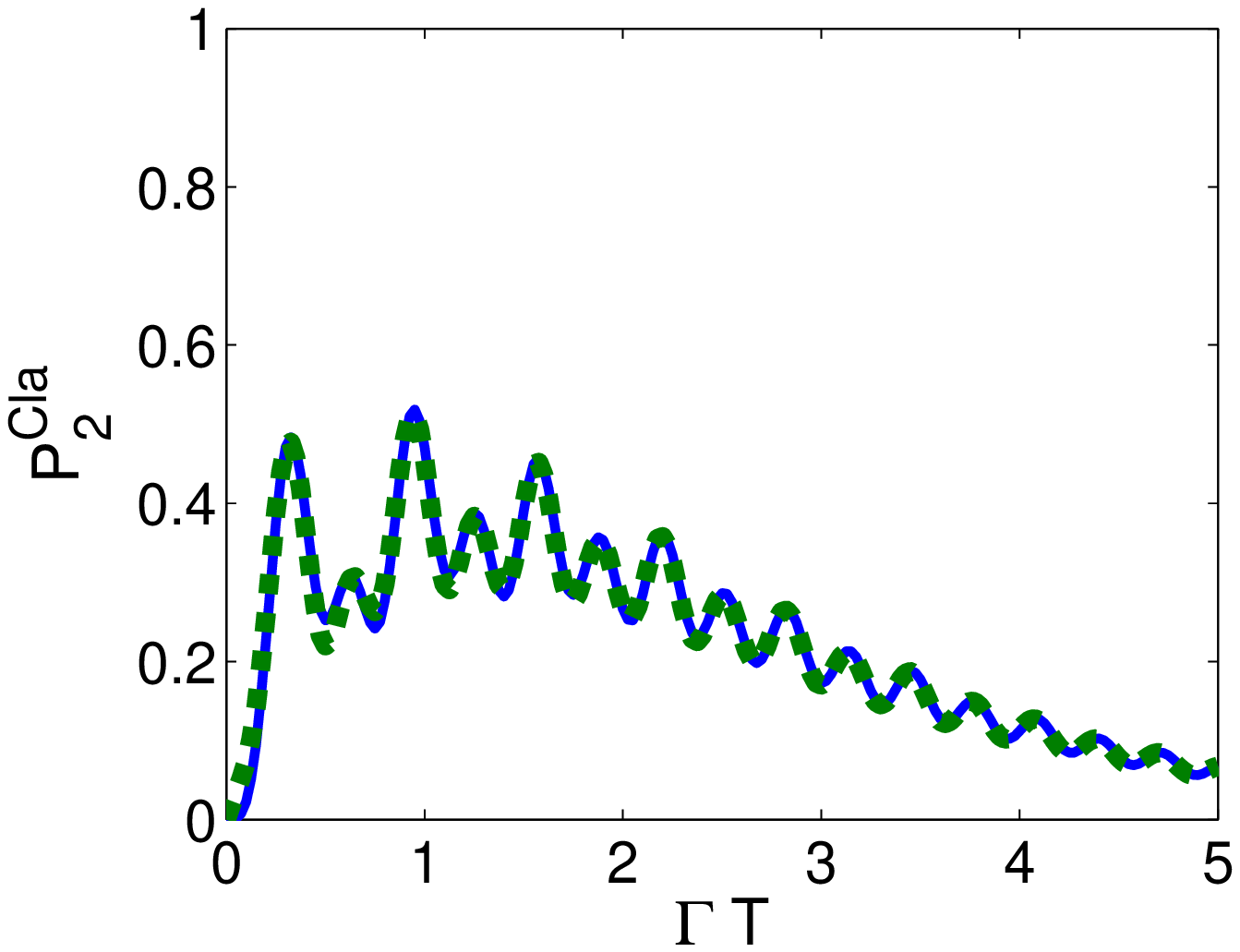}\textbf{f}\\
\end{eqnarray*}
\caption{The semiclassical parts of probabilities of zero, one and
two (from left to right) photon emissions from the two laser pulses
as a function of T - the duration of a pulse with $\Omega=10\Gamma$.
The solid line is the exact result Eqs.~(\ref{P0exact}), (\ref{P1})
and (\ref{P2}) and the dashed curve show the approximation in the
limit of strong fields Eqs.~(\ref{eqppzz}), (\ref{P1exact}) and
(\ref{P2exact}). The upper row (\textbf{a}, \textbf{b}, \textbf{c})
illustrates the result for $\Delta=3$ and the lower row (\textbf{e},
\textbf{f}, \textbf{d}) illustrates the same probabilities for
$\Delta=0.5$.}
\end{figure}
\label{Correl}
\end{widetext}
 \subsection{Strong and Short Pulses}

   The sequence of two very short and strong pulses is
important due to its numerous practical applications. Mathematically
we define this limit as $T\rightarrow0$, $\Omega\rightarrow\infty$
in such a way, that the product $\Omega T$ stays of the order of
unity $\Omega T\sim1$ (automatically leading also to
$\Omega\gg\Gamma$), otherwise the molecule will never reach the
excited state, and the probability to obtain non-zero results for
$P_1$ and $P_2$ will be negligible. As a consequence, the
spontaneous emission process during the pulse event may be
neglected, and therefore, it is reasonable to approximate the
behavior of the system by simple Shr\"{o}dinger evolution with
well-known Rabi oscillations. In this limit the photons can be
emitted only in the delay interval or after the second pulse, while
the only non-zero propagator acting on the molecule during the
pulses within RWA has the following matrix representation :
\begin{equation}
\lim_{\begin{array}{c}
_{T\rightarrow0,}\\
_{\Omega\rightarrow\infty}\end{array}} \tilde{U}^{(0)}_{(T,0)} =
\left(
\begin{array}{c c c c}
  \cos^2 {\Omega T \over 2}  &  \sin^2 { \Omega T \over 2}   & - i {\sin \Omega T  \over 2}  & i {\sin \Omega T \over 2} \\
 \sin^2 { \Omega T \over 2}  & \cos^2 { \Omega T \over 2}   &  i {\sin \Omega T  \over 2}  & - i  {\sin \Omega T \over 2} \\
 -i {\sin \Omega T \over 2} & i {\sin \Omega T \over 2}  &  \cos^2 {\Omega T \over 2}  &  \sin^2 { \Omega T \over 2}  \\
 i {\sin \Omega T \over 2} & - i  {\sin \Omega T \over 2}  &  \sin^2 {\Omega T \over 2} &  \cos^2 { \Omega T \over 2}
\end{array}
\right), \label{RabiOscl}
\end{equation}
which is independent of $\Gamma$. Note, that since this
zero-photon-propagator Eq.~(\ref{RabiOscl}) describes the
conservative evolution of the system, the transformation is unitary,
all the elements exhibit Rabi oscillations, and symmetry and
reversibility of the matrix elements are found: $\langle {\rm
e}|\tilde{U}^{(0)}_{(T,0)}|{\rm e}\rangle=\langle {\rm g} |
\tilde{U}^{(0)}_{(T,0)}| g\rangle$, $\langle {\rm e}
|\tilde{U}^{(0)}_{(T,0)}|{\rm g}\rangle=\langle{\rm g}
|\tilde{U}^{(0)}_{(T,0)}|{\rm e}\rangle$ etc.\\

  Now we consider $P_0$, $P_1$ and $P_2$ in the limit of the short and strong
pulses and demonstrate how this physical constrain can help in
reducing the number of paths appearing in
Eqs.~(\ref{eqPn1}),(\ref{Pncla}),(\ref{Ancoh}). The exact
expressions for $U^{(1)}_{(t,0)}$ and $U^{(2)}_{(t,0)}$, according
to Eq.~(\ref{eqPngen}) consist of the sum of 4 and 10 terms
respectively - see Eqs.~(\ref{eqU1exact}), (\ref{eq2exact}) in
Appendix C. After neglecting the trajectories where photons are
emitted within the pulse events (since $T\to0)$) we have :
$$\lim_{T\to0,\Omega\to\infty}U^{(1)}_{(t,0)}=U^{(1)}_{(t,t_3)}U^{(0)}_{(t_3,t_2)}U^{(0)}_{(t_2,t_1)}U^{(0)}_{(t_1,0)}$$
\begin{equation}
+U^{(0)}_{(t,t_3)}U^{(0)}_{(t_3,t_2)}U^{(1)}_{(t_2,t_1)}U^{(0)}_{(t_1,0)}
\label{U1shortpulses}
\end{equation}
and
\begin{equation}
\lim_{T\to0,\Omega\to\infty}U^{(2)}_{(t,0)}=U^{(1)}_{(t,t_3)}U^{(0)}_{(t_3,t_2)}U^{(1)}_{(t_2,t_1)}U^{(0)}_{(t_1,0)}.
\label{U2shortpulses}
\end{equation}
This is one example where the formulation of photon statistics based
on quantum trajectories is very convenient, since we can identify
the underlying physical processes and make approximations. Inserting
the closure relation Eq.~(\ref{closure}) between every two
propagators of Eqs.~(\ref{U1shortpulses}), (\ref{U2shortpulses}) and
using the matrix elements of the zero-photon-propagator
Eq.~(\ref{RabiOscl}), we obtain the leading semiclassical and
coherence terms of $P_0$, $P_1$ and $P_2$ in the limit of the short
and strong pulses. The results are summarized in Table 1 below and
they are valid for any sequence of short pulses.
\begin{widetext}
$$ \mbox{
\begin{tabular}{|c| c| c|}
\hline
$n$ & $p_n^{{\rm Cla}}$ & $a_n ^{{\rm Coh}}$ \\
\hline $ 0$ &  $ \langle {\rm g} | \tilde U^{(0)}_{(t_3,t_2)} | {\rm
g} \rangle\langle {\rm g} | \tilde U^{(0)}_{(t_1,0)}| {\rm g}
\rangle + e^{-\Gamma\Delta}\langle {\rm g} | \tilde
U^{(0)}_{(t_3,t_2)}| {\rm e} \rangle \langle {\rm e} |\tilde
U^{(0)}_{(t_1,0)}| {\rm g} \rangle$&
$ \langle {\rm g} | \tilde U^{(0)}_{(t_3,t_2)}|c\rangle \langle c | \tilde U^{(0)}_{(t_1,0)}| {\rm g} \rangle $ \\
\ & \ & \\
\hline $ 1$ &  $ \langle {\rm g} | \tilde U^{(0)}_{(t_3,t_2)} | {\rm
g} \rangle\langle {\rm e} | \tilde U^{(0)}_{(t_1,0)}| {\rm g}
\rangle + \langle {\rm e} | \tilde U^{(0)}_{(t_3,t_2)}| {\rm g}
\rangle \langle {\rm g} |\tilde U^{(0)}_{(t_1,0)}| {\rm g} \rangle$&
$ \langle {\rm e} | \tilde U^{(0)}_{(t_3,t_2)}|c\rangle \langle c | \tilde U^{(0)}_{(t_1,0)}| {\rm g} \rangle $ \\
\ & \ & \\
\hline $ 2$ & $ \langle {\rm e} |\tilde U^{(0)}_{(t_3,t_2)} | {\rm
g} \rangle \langle {\rm e} |\tilde U^{(0)}_{(t_1,0)}| {\rm g}
\rangle\left(1 - e^{ - \Gamma \Delta} \right) $ &
$ 0 $ \\
& \ & \  \\
\hline
\end{tabular}}$$
$$\mbox{\textbf{Table 1:} Photon statistics for short pulses.}$$
Note, that the coherence terms of $P_2$ vanish, since emission of a
photon in the delay interval, necessary for emitting two photons in
the limit of very short pulses, destroys the coherence . Using the
symmetry and reversibility of the zero-photon-propagator matrix
elements Eq.~(\ref{RabiOscl}), it is easy to show that $a_{0}^{\rm
Coh}=-a_{1}^{\rm Coh}$ and $ p_0^{\rm Cla}+ p_1 ^{\rm Cla} + p_2
^{\rm Cla}=1$, i.e. the semiclassical paths conserve probability.
Finally, we bring attention to the fact, that the semiclassical
paths of $P_1$ do not depend neither on the spontaneous emission
rate $\Gamma$ not on the delay interval duration $\Delta$ (see
Fig.~2.{\bf b} and Fig.~2.{\bf e}). Comparing the two non-negligible
trajectories of $p_1^{\rm Cla}$ with Eq.~(\ref{Pncorrel}), we see
that they correspond to the first term of the righthand side of this
equation - i.e to the product of probabilities of emitting 0 and 1
photons, related to each one of the pulses independently.\\

Using the matrix elements  Eq.~(\ref{RabiOscl}) and Table 1, after
some algebra we obtain explicitly :
\begin{equation}
\lim_{T\to0,\Omega\to\infty}P_0^{\rm} =
e^{-\Delta}\sin^4\left({\Omega T \over 2}\right) + \cos^4
\left({\Omega T \over 2}\right)- {1\over 2}  e^{-\Delta/2} \sin^2
\left(\Omega T\right)
\cos\left[\omega_0\left(T+\Delta\right)-\phi_2\right]
\label{standshP0}
\end{equation}
which is the $\Omega\rightarrow\infty$, $T \to 0$ limit of
Eq.~(\ref{P0exact}).
\begin{equation}
\lim_{T\to0,\Omega\to\infty}P_1^{\rm } = {1 \over 2} \sin^2 (\Omega
T )\left\{ 1  +  e^{ - {\Delta/2} }  \cos\left[ \omega_0 \left( T +
\Delta\right)-\phi_2 \right]\right\} \label{Eqpooo}
\end{equation}
\end{widetext}
and
\begin{equation}
\lim_{T\to0,\Omega\to\infty}P_2^{\rm}= \left(1 - e^{ -
\Delta}\right) \sin^4 \left({\Omega T \over 2}\right) \label{Eqpttt}
\end{equation}
It is easy to see, that when $T\rightarrow0$ the
Eqs.~(\ref{P1exact}), (\ref{P2exact}) reduce to Eqs.~(\ref{Eqpooo}),
(\ref{Eqpttt}) as expected.\\

For a $\pi$-pulse defined by $\Omega T = \pi + 2 \pi n$
\cite{Yong,Yong1}, where $n$ is a non negative integer, in the
strong field limit we obtain :
\begin{equation}
\lim_{T\to0,\Omega\to\infty}P_0 \sim e^{ - \Delta}.
\end{equation}
This behavior may be easily understood for : substituting $\Omega T
= \pi$ into Eq.~(\ref{RabiOscl}) we have
\begin{equation}
\lim_{\begin{array}{c}
_{T\rightarrow0,\Omega\rightarrow\infty},\\_{\Omega
T=\pi}\end{array}} \tilde{U}^{(0)}_{(T,0)} = \left(
\begin{array}{c c c c}
0 & 1 & 0 & 0 \\
1 & 0 & 0 & 0\\
0 & 0 & 0 & 1 \\
0 & 0 & 1 & 0
\end{array}
\right) \label{Pipulse}
\end{equation}
The physical meaning of this propagator follows directly from its
matrix representation : as well-known, the ideal $\pi$-pulse simply
switches the state of the molecule from the excited to the ground
state and vice versa. Thus, the first $\pi$-pulse of the sequence
pumps the molecule from the ground state to the excited state. If
the delay between the pulses $\Delta$ is long, the molecule will
emit a photon before the arrival of the second pulse, and then
$\lim_{T\to0,\Omega\to\infty}P_0 = 0$. Contrary, if $\Delta\ll 1$,
the second $\pi$-pulse pushes the molecule back to the ground state
before the emission of a photon, and then
$\lim_{T\to0,\Omega\to\infty}P_0 = 1$. We also notice, that the
interference $\cos \left[ \omega_0 \left( T + \Delta \right)-\phi_2
\right]$ term vanishes for the $\pi$-pulses, since according to
Eq.~(\ref{Pipulse}) the off-diagonal matrix elements giving raise to
this term are equal zero. On the opposite, for a $\pi/2$-pulse,
defined by $\Omega T = \pi/2 + 2 \pi n $, the influence of the
coherence on the photon statistics generally does not vanish:
\begin{equation}
\lim_{\Omega\gg1}P_0 \sim {1 \over 4} \left\{ e^{ - \Delta} + 1 - 2
e^{- \Delta/2} \cos\left[ \omega_0(\Delta+T)\right] \right\}
\end{equation}
Once again we remind, that in optics in many cases the ideal $\pi$
and $\pi/2$-pulses are considered where the interaction time $\Gamma
T \to 0$, and then  $e^{ - T} = 1$, but $\omega_0 T$ is not a small
number, especially because we work under the assumption
$\Omega\ll\omega_0$, which is essential for the two level model
approximation of the molecule and for the assumption, that the
spontaneous emission rate $\Gamma$ is not effected by the presence
of the
laser field \cite{CT}.\\

\subsection{Weak and Long Pulses}

  Here we consider the case of very long and weak pulses. In this limit the delay interval
has a negligible effect on $P_n$. In this limit we expect, that
according to Eq.~(\ref{Pncorrel}) the probability of emitting n
photons from two separated pulses may be approximated by
$P_n^{I_2I_1}$ - the probability of emitting n photons from two
attached pulses with zero delay :
\begin{equation}
\lim_{T\rightarrow\infty}P_n=P_n^{I_2I_1}=\sum_{\alpha=0}^n\langle
g|U^{(\alpha)}_{(2T)}|g\rangle + \sum_{\alpha=0}^{n-1}\langle
e|U^{(\alpha)}_{(2T)}|g\rangle \label{LongT}
\end{equation}
 Taking the limit
$\Omega\rightarrow0$, $T\rightarrow\infty$ in such a way that
 $\Omega^{2}T$ remains finite of the
  exact solution for $P_0$, $P_1$ and $P_2$ Eqs.~(\ref{P0exact}),(\ref{P1}),(\ref{P2}), we find
\begin{equation}
\lim_{\Omega\to0,T\to\infty}P_0 = e^{-2 \Omega^2 T }\label{P0weak}
\end{equation}

\begin{equation}
\lim_{\Omega\to0,T\to\infty}P_1 = 2 \Omega^2 T e^{-2 \Omega^2 T }
\label{P1weak}
\end{equation}
\begin{equation}
\lim_{\Omega\to0,T\to\infty}P_2 =  {(2 \Omega^2 T )^2 \over 2! }
e^{-2 \Omega^2 T }\label{P2weak}
\end{equation}
\begin{equation}
\lim_{\Omega\to0,T\to\infty}P_n = {( 2 \Omega^2 T )^n  \over n! }
e^{-2 \Omega^2 T } \label{Poissonian}
\end{equation}\\
Indeed one can show, that Eq.~(\ref{LongT}) and the limiting
behavior Eqs.(\ref{P0weak}-\ref{Poissonian}) of the exact results
are identical. Note, that Eq.~(\ref{Poissonian}) corresponds to
Poissonian statistics. This behavior originates from the fact, that
because of the long pulses duration and weak laser field, the
leading terms of $P_n$ are those, where photon emissions are well
separated one from another on the time axis, and therefore photon
statistics is described by nearly uncorrelated emission events.
\subsection{ The upper and lower bounds for strong pulses}

  Finally we investigate the upper and lower bounds of $P_0$, $P_1$ and $P_2$
within the strong field approximation. Calculating the first and the
second order partial derivatives with respect to T, we find the
extremum  of  Eqs.~(\ref{eqppzz}), (\ref{P1exact}) and
(\ref{P2exact}), and neglecting the ultra-fast oscillating coherence
terms, obtain the bounds of semiclassical parts of $P_0$, $P_1$ and
$P_2$ :
\begin{equation}
  e^{-(T+\Delta)} \leq p_0^{\rm Cla}\leq  e^{-T},
  \label{boundP0}
\end{equation}
\begin{equation}
 {1 \over 4} T e^{-(T+\Delta)} \leq p_1^{\rm Cla} \leq  {e^{-T} \over 8} \left[ 4 + 3 T \left( 1 + e^{-\Delta} \right )
 \right],
\label{boundP1}
\end{equation}
\begin{equation}
  { 3 \over 4} T e^{-T} \leq p_2^{\rm Cla} \leq  e^{-T} \left( 1 - e^{-\Delta} +  { 3 \over 4} T
  \right).
\label{boundP2}
\end{equation}
where once again $\Gamma=1$.\\

  The origin of  Eqs.~(\ref{boundP0}),(\ref{boundP1}) and
(\ref{boundP2}), although non-trivial for a finite T, can be easily
understood in the limit of the short pulses $T\to0$. The upper bound
of $p_0^{\rm Cla}$ is obvious, since if the interaction time is
zero, no photons will be emitted for sure. Further, since we neglect
the probability of emitting photons during the pulses, $p_0^{\rm
Cla}$ may only be decreased by the probability of not emitting a
photon during the delay interval - $e^{-\Delta}$. Considering
$p_1^{\rm Cla}$ we see, that Eq.~(\ref{Eqpooo}) reaches the maximum
value - $\frac{1}{2}$, when $\Omega T=\pi/2$. Hence, the
maximization of $p_1^{\rm Cla}$ corresponding to the interaction
with a sequence of two pulses is achieved by applying two ideal
$\pi/2$-pulses. Finally, from Eq.~(\ref{Eqpttt}) for
$\lim_{T\to0,\Omega\to\infty}P_2$ follows, that maximum of this
expression is found for $\Omega T=\pi$, and hence the optimization
of
$P_2$ is achieved by two ideal $\pi$-pulses.\\

  Learning from this simple
example, although not rigorously, we extend it to the conclusion,
that the maximum of $P_n$ ($n>1$) for any fixed interaction time is
optimized by producing n equally separated $\pi$-pulses. This
statement follows from the following argument : as far as we work
under assumption, that the laser field does not effect the
spontaneous emission process, the maximization of $P_n$ in any
limited time interval may be achieved by minimization of induced
emission, which is guaranteed by the
strong and short ideal $\pi$-pulses better than any by any others.\\

   In Fig.~3 we show the maximum of $p_2^{\rm Cla}$ from a sequence
of two equal square pulses as a function of the interaction's
strength $\Omega$ and pulse's duration $T$ for three fixed values of
the total interaction time $2T+\Delta$ . The curves were obtained
using the extremum conditions of the exact expression for $P_2$
Eq.~(\ref{P2}). From Fig.~3 we see, that as the interaction time
$\Gamma T$ becomes shorter, the delay period $\Delta$ longer and the
Rabi frequency $\Omega$ larger, the probability of emitting 2
photons is getting close to 1. Although it becomes equal exactly 1
only for two ideal $\pi$-pulses separated by infinite delay, the
graph shows, that starting from some range of parameters the
increasing of $p_2^{\rm Cla}$ slows down, so that further increasing
$\Omega$ does not contribute much. Finally, we note that for short
delay intervals $\Delta<1$ the maximum of $p_2^{\rm Cla}$
is much less than 1 as expected.\\
\begin{figure}
\includegraphics[width=\columnwidth]{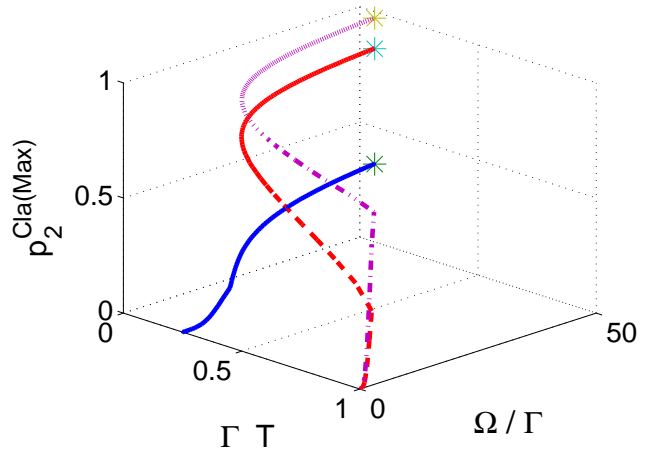}\\
\caption{The maximum of the probability of emission of two photons
in a two-pulse laser field. The $t_3$ of the end time of second
pulse is fixed at $0.5$ (solid curve), $2.0$ (dashed curve), $4.5$
(dot-dashed curve), respectively.  The delay time between two
pulses, $\Delta= t_3-2T$. The star gives the asymptotic behavior of
Eq.~(\ref{P2}) in the limit of strong fields ($\Omega$ =50) and
short pulse with $\Omega T = \pi$, $\Gamma=1$.}\label{p2max}
\end{figure}
\section{Summary}

  We obtained general expressions for the probability of n photon emission events
for a two level system interacting with two laser pulses separated
by a delay interval $\Delta$. The photon statistic was represented
as summation over quantum trajectories, which allowed simple
intuitive physical interpretation of the final results
Eqs.~(\ref{eqPn1}), (\ref{Pncla}), (\ref{Ancoh}). In particular, the
contribution of the coherence effect, resulting from the quantum
trajectories going through the superposition of the pure states at
the end of the first pulse, was discussed. Although in optics it
might be difficult to detect this effect experimentally, since the
coherence paths oscillate in $\Delta$ with extremely large molecule
absorption frequency $\omega_0$, nevertheless in microwave
spectroscopy, dealing with lower range of absorption frequencies,
the coherence effect is important \cite{Katz}. In addition, the
correlation function $C(\Delta)$ Eq.~(\ref{correl}) was suggested as
a measure of the photon statistics deviation from a treatment where
the sequence of pulses is considered as if the pulses were
independent. This correlation might be a useful tool to quantify the
coherence and ``memory" of single molecules, atoms or quantum dots
through the measured photon statistics. \\

  The application of our general results was
demonstrated on detailed calculation of $P_0$, $P_1$ and $P_2$ - the
probabilities of emitting 0, 1 and 2 photons from the sequence of
two square laser pulses Eqs.~(\ref{P0exact}), (\ref{P1}) and
(\ref{P2}) in the limit of long measurement times $t\to\infty$. The
physical interpretation of the quantum paths was shown to be useful
in reducing the complexity of calculations in the limit of short and
strong pulses (e.g. neglecting the paths where the photons are
emitted within the pulse events). Finally, the non-trivial upper and
lower bounds for the strong square pulses with finite duration were
obtained. This kind of information is useful in experiments, where
pulses are neither infinitely short not infinitely strong. Our
approach can be applied to the theoretical study of other types of
non-linear spectroscopy such as three level systems, systems
undergoing stochastic dynamics \cite{SB}, or Josephson junction
qubits \cite{Katz} controlled by microwave radiation, where the
strong dependence on the contribution of coherence was already experimentally proved.\\

{\bf Acknowledgment:} This work was supported by the Israel Science
Foundation.\\
\begin{widetext}
\section{Appendix A}
In this appendix we give the detailed derivation of
Eq.~(\ref{Pncorrel}).  Consider the semiclassical trajectories
Eq.~(\ref{Pncla}) of $P_n$ from the two pulses . First let us sort
the terms according to
$$P_n^{\rm Cla}=\sum_{\alpha=0}^{n}\langle g|U^{(n-\alpha)}_{(t_3,t_2)}|g\rangle
\langle
g|U^{(\alpha)}_{(t_1,0)}|g\rangle+\sum_{\alpha=0}^{n-1}\langle
e|U^{(n-\alpha-1)}_{(t_3,t_2)}|g\rangle \langle
g|U^{(\alpha)}_{(t_1,0)}|g\rangle +\sum_{\alpha=0}^{n-1}\langle
g|U^{(n-\alpha-1)}_{(t_3,t_2)}|g\rangle \langle
e|U^{(\alpha)}_{(t_1,0)}|g\rangle$$ $$ +\sum_{\alpha=0}^{n-2}\langle
e|U^{(n-\alpha-2)}_{(t_3,t_2)}|g\rangle \langle
e|U^{(\alpha)}_{(t_1,0)}|g\rangle +e^{-\Delta\Gamma}\left\{
\sum_{\alpha=0}^{n}\langle g|U^{(n-\alpha)}_{(t_3,t_2)}|e\rangle
\langle e|U^{(\alpha)}_{(t_1,0)}|g\rangle+
\sum_{\alpha=0}^{n-1}\langle e|U^{(n-\alpha-1)}_{(t_3,t_2)}|e\rangle
\langle e|U^{(\alpha)}_{(t_1,0)}|g\rangle\right.$$
\begin{equation}\left. -\sum_{\alpha=0}^{n-1}\langle
g|U^{(n-\alpha-1)}_{(t_3,t_2)}|g\rangle \langle
e|U^{(\alpha)}_{(t_1,0)}|g\rangle-\sum_{\alpha=0}^{n-2}\langle
e|U^{(n-\alpha-2)}_{(t_3,t_2)}|g\rangle \langle
e|U^{(\alpha)}_{(t_1,0)}|g\rangle \right\}
\label{Pncla0.1}\end{equation}
Now we consider the sum of the first two terms
$$\sum_{\alpha=0}^{n}\langle g|U^{(n-\alpha)}_{(t_3,t_2)}|g\rangle
\langle
g|U^{(\alpha)}_{(t_1,0)}|g\rangle+\sum_{\alpha=0}^{n-1}\langle
e|U^{(n-\alpha-1)}_{(t_3,t_2)}|g\rangle \langle
g|U^{(\alpha)}_{(t_1,0)}|g\rangle$$
$$ =\left\{\langle
g|U^{(n)}_{(t_3,t_2)}|g\rangle +\langle
e|U^{(n-1)}_{(t_3,t_2)}|g\rangle\right\} \langle
g|U^{(0)}_{(t_1,0)}|g\rangle+\left\{\langle
g|U^{(n-1)}_{(t_3,t_2)}|g\rangle +\langle
e|U^{(n-2)}_{(t_3,t_2)}|g\rangle\right\} \langle g|U^{(1}_{(t_1,0)}|g\rangle+\cdots$$\\
\begin{equation}
=P_n^{I_2}\langle g|U^{(0)}_{(t_1,0)}|g\rangle+ P_{n-1}^{I_2}\langle
g|U^{(1)}_{(t_1,0)}|g\rangle+\cdots=\sum_{\alpha=0}^n
P_{n-\alpha}^{I_2}\langle g|U^{(\alpha)}_{(t_1,0)}|g\rangle,
\label{Pncla0.2}
\end{equation}
Similar manipulations with the second two terms lead to :
\begin{equation} \sum_{\alpha=0}^{n-1}\langle
g|U^{(n-\alpha-1)}_{(t_3,t_2)}|g\rangle \langle
e|U^{(\alpha)}_{(t_1,0)}|g\rangle+\sum_{\alpha=0}^{n-2}\langle
e|U^{(n-\alpha-2)}_{(t_3,t_2)}|g\rangle \langle
e|U^{(\alpha)}_{(t_1,0)}|g\rangle = \sum_{\alpha=0}^{n-1}
P_{n-1-\alpha}^{I_2}\langle e|U^{(\alpha)}_{(t_1,0)}|g\rangle
\label{Pncla0.3}
\end{equation}
Combining Eqs.~(\ref{Pncla0.2}) and (\ref{Pncla0.3}) we get :
$$
\sum_{\alpha=0}^n P_{n-\alpha}^{I_2}\langle
g|U^{(\alpha)}_{(t_1,0)}|g\rangle+\sum_{\alpha=0}^{n-1}
P_{n-1-\alpha}^{I_2}\langle e|U^{(\alpha)}_{(t_1,0)}|g\rangle=
$$
\begin{equation}
P_{n}^{I_2}\langle
g|U^{(0)}_{(t_1,0)}|g\rangle+\left\{P_{n-1}^{I_2}\langle
e|U^{(0)}_{(t_1,0)}|g\rangle+P_{n-1}^{I_2}\langle
g|U^{(1)}_{(t_1,0)}|g\rangle\right\}+\left\{P_{n-2}^{I_2}\langle
e|U^{(1)}_{(t_1,0)}|g\rangle+P_{n-2}^{I_2}\langle
g|U^{(2)}_{(t_1,0)}|g\rangle\right\}+\cdots =\sum_{\alpha=0}^n
P_{n-\alpha}^{I_2}P_{\alpha}^{I_1} \label{Pncla0.4}
\end{equation}
Now we concentrate on the terms multiplied by the factor
$e^{-\Gamma\Delta}$ in Eq.~(\ref{Pncla0.1}). Let's add and subtract
the two following terms :$$\sum_{\alpha=0}^{n}\langle
g|U^{(n-\alpha)}_{(t_3,t_2)}|g\rangle \langle
g|U^{(\alpha)}_{(t_1,0)}|g\rangle+\sum_{\alpha=0}^{n-1}\langle
e|U^{(n-\alpha-1)}_{(t_3,t_2)}|g\rangle \langle
g|U^{(\alpha)}_{(t_1,0)}|g\rangle.$$
 We get
$$
e^{-\Delta\Gamma}\left\{\sum_{\alpha=0}^{n}\langle
g|U^{(n-\alpha)}_{(t_3,t_2)}|e\rangle \langle
e|U^{(\alpha)}_{(t_1,0)}|g\rangle+\sum_{\alpha=0}^{n-1}\langle
e|U^{(n-\alpha-1)}_{(t_3,t_2)}|e\rangle \langle
e|U^{(\alpha)}_{(t_1,0)}|g\rangle+ \sum_{\alpha=0}^{n}\langle
g|U^{(n-\alpha)}_{(t_3,t_2)}|g\rangle \langle
g|U^{(\alpha)}_{(t_1,0)}|g\rangle\right.$$
$$+\sum_{\alpha=0}^{n-1}\langle
e|U^{(n-\alpha-1)}_{(t_3,t_2)}|g\rangle \langle
g|U^{(\alpha)}_{(t_1,0)}|g\rangle -\sum_{\alpha=0}^{n}\langle
g|U^{(n-\alpha)}_{(t_3,t_2)}|g\rangle \langle
g|U^{(\alpha)}_{(t_1,0)}|g\rangle
 -\sum_{\alpha=0}^{n-1}\langle
e|U^{(n-\alpha-1)}_{(t_3,t_2)}|g\rangle \langle
g|U^{(\alpha)}_{(t_1,0)}|g\rangle$$
\begin{equation}\left. -\sum_{\alpha=0}^{n-1}\langle
g|U^{(n-\alpha-1)}_{(t_3,t_2)}|g\rangle \langle
e|U^{(\alpha)}_{(t_1,0)}|g\rangle-\sum_{\alpha=0}^{n-2}\langle
e|U^{(n-\alpha-2)}_{(t_3,t_2)}|g\rangle \langle
e|U^{(\alpha)}_{(t_1,0)}|g\rangle\right\}
\label{Pncla0.5}\end{equation} But the first 4 paths of
Eq.~(\ref{Pncla0.5}) are just the semiclassical part of probability
of emitting n photons from the two pulses attached together
\begin{equation}
P_n^{Cla,I_2I_1}=\sum_{\alpha=0}^n\langle
g|U^{(n-\alpha)}_{(t_3,t_2)} \left(|g \rangle \langle g|+ |e \rangle
\langle e| \right) U^{(\alpha)}_{(t_1,0)}|g\rangle +
\sum_{\alpha=0}^{n-1}\langle e|U^{(n-1-\alpha)}_{(t_3,t_2)}\left(|g
\rangle \langle g|+ |e \rangle \langle e| \right)
U^{(\alpha)}_{(t_1,0)}|g\rangle \label{pi2i1cla} \end{equation}
(compare with Eq.~(\ref{eqPnt1})). And the last 4 paths of
Eq.~(\ref{Pncla0.5}) are equal to
Eq.~(\ref{Pncla0.2})+Eq.~(\ref{Pncla0.3}). Putting all this
information together we obtain
\begin{equation}
P_n=\sum_{\alpha=0}^{n}P_{n-\alpha}^{I_2}P_{\alpha}^{I_1}+
e^{-\Delta\Gamma}\left\{P_{n}^{Cla,
I_2I_1}-\sum_{k=0}^{n}P_{n-\alpha}^{I_2}P_{\alpha}^{I_1}+
(e^{\Delta(\Gamma/2+i\omega_0)}A_n^{\rm Coh}+C.C.)\right\}
\label{Pncla0.6}
\end{equation}
Finally, by addition and substraction of the coherence trajectories:
$$e^{-\Delta\Gamma}\left\{A_n^{Coh,I_2I_1}+C.C.\right\}=e^{-\Delta\Gamma}\left\{\sum_{\alpha=0}^n\langle
g|U^{(n-\alpha)}_{(t_3,t_2)} \left(|c \rangle \langle c|+ |c^*
\rangle \langle c^*| \right) U^{(\alpha)}_{(t_1,0)}|g\rangle +
\sum_{\alpha=0}^{n-1}\langle e|U^{(n-1-\alpha)}_{(t_3,t_2)}\left(|c
\rangle \langle c|+ |c^*\rangle \langle c^*| \right)
U^{(\alpha)}_{(t_1,0)}|g\rangle\right\}$$
$$=e^{-\Delta\Gamma}\left\{A_n^{\rm Coh}+C.C.\right\}$$
to $P_{n}^{Cla,I_2I_1}$ we obtain Eq.~(\ref{Pncorrel}). Using
Eq.~(\ref{Pncorrel}) it should be taken into account however that
not all the coherence paths now oscillate in $\Delta$ with
$\omega_0$.
\end{widetext}

\section{Appendix B}
   The Rotating Wave Approximation (RWA) \cite{CT} consists of
neglecting the non-resonant processes of rising from $|g\rangle$ to
$|e\rangle$ by emitting a photon and falling from $|e\rangle$ to
$|g\rangle$ by absorbing a photon. Switching to the rotating frame
by applying the transformation $A_{(t-t',\phi,\omega_L)}$ defined
below, it is possible to suppress any time dependence in the Bloch
equation Eq.~(\ref{eqBloch}). As a result the following time
independent equation is obtained
\begin{equation}
\dot{\tilde{\sigma}}_{(t)} = \left[\tilde L+
\hat{\Gamma}\right]\tilde{\sigma}_{(t)}\label{eqBloch1}
\end{equation}
where
\begin{equation}\tilde{\sigma}_{(t)}=A_{(t-t',\phi,\omega_L)}\sigma_{(t)},
\label{tildsig}
\end{equation}
$t'$ - is the initial moment,
\begin{equation}
\tilde L = \left(
\begin{array}{c c c c}
-\Gamma  & 0 & {-i \Omega \over 2} & {i \Omega \over 2} \\
0 & 0 & {i \Omega \over 2} & {-i \Omega \over 2} \\
 {-i \Omega \over 2} & {i \Omega \over 2} & - {\Gamma\over 2} - i \delta_L & 0 \\
 {i \Omega \over 2} & {-i \Omega \over 2} & 0 & - {\Gamma\over 2} + i \delta_L
\end{array}
\right) \label{eqLRWA}
\end{equation}
and the transformation $A_{(t-t',\phi,\omega_L)}$ is given by
\begin{equation}
A_{(t-t',\phi,\omega_L)} = \left(
\begin{array}{c c c c}
1 & 0 & 0  & 0 \\
0 & 1 & 0 & 0 \\
0 & 0 & e^{-i (\omega_L(t-t') + \phi)} & 0 \\
0 & 0 & 0 & e^{i (\omega_L(t-t') + \phi)}
\end{array}
\right), \label{eqA}
\end{equation}
where $\phi$ is the phase of the laser at the initial moment $t'$.
 In the new representation the calculation of the Green function
is straightforward (see Eq.~(\ref{eqtildeGreenf})). Representing the
solution to the time independent Bloch equation Eq.~(\ref{eqBloch1})
in the rotating frame as the infinite iterative expansion in $\hat
\Gamma $ we find the following expression for n-photon-propagator
within RWA
\begin{equation}
\tilde{U}^{(n)}_{(t-t')}=\int_{t'} ^t  \cdots \int_{t'} ^{t_2}
\tilde{{\cal G}}(t-t_n) \hat{\Gamma} \cdots \hat{\Gamma}
\tilde{{\cal G}}(t_{1}-t') {\rm d} t_1 \cdots {\rm d} t_n.
\label{eqtildUn}
\end{equation}
Hence
\begin{equation}
\tilde{\sigma}^{(n)}_{(t)}=\tilde{U}^{(n)}_{(t-t')}\tilde{\sigma}_{(t')}.
\label{sigmaRWA}
\end{equation}
where $\tilde{\sigma}_{(t')}=A_{(0,\phi,\omega_L)}\sigma_{(t')}$ is
the initial condition. Finally for obtaining $\sigma^{(n)}_{(t)}$ we
have to apply the inverse transformation
\begin{equation}
\sigma^{(n)}_{(t)}= A^{-1}_{(t-t',\phi,\omega_L)}\tilde{U}^{
(n)}_{(t-t')}A_{(0,\phi,\omega_L)}\sigma_{(t')} \label{sigRWA}
\end{equation}
From Eq.~(\ref{sigRWA}) one easily makes the following conclusions:
(i) the initial state of the molecule is multiplied by
$A_{(0,\phi_1,\omega_L)}$, where $\phi_1$ is the initial phase of
the laser at the beginning of the first pulse, which shifts the
initial coherence phase by $-\phi_1$ (ii) the delay period
propagators are now multiplied by $A_{(0,\phi_2,\omega_L)}$ from the
left due to the second pulse and by $A^{-1}_{(T,\phi_1,\omega_L)}$
from the right due to the first pulse ($\phi_2$ is the initial laser
phase at the beginning of the second pulse). Clearly, this leads
only to an additional phase shift $(T\omega_L +\phi_1-\phi_2)$ of
the coherence terms. Therefore calculating the photon statistics for
the square pulses we rewrite the
 Eqs.~(\ref{eqPn1}),(\ref{Pncla}),(\ref{Ancoh}) with the following modifications:\\

  1) In the definition of the n-photon-propagator Eq.~(\ref{eqUn})
the Green function defined as the time ordered exponential are
replaced by
\begin{equation}
\tilde{{\cal G}} (t-t')= \exp\left[(t-t') \tilde L
\right]\label{eqtildeGreenf}
\end{equation}
which are Green functions for the time intervals inside the pulses
within RWA.\\

2) The initial state of the system must be replaced by
$A_{(0,\phi,\omega_L)}\sigma_{(0)} $\\

3) The coherent terms $A_n^{\rm Coh}$ are multiplied by additional
phase factor
 $e^{i(\omega_L T+\phi_1-\phi_2)}$.\\
\textit{Remark : If in experiments the initial phases of the pulses
are random variables, it's necessary to replace all the phase
factors with their ensemble averages.} \\

Thus summarizing we have
\begin{equation} P_n=p_{n}^{\rm Cla}+
e^{-\Delta\Gamma/2}\left[e^{i
(\Delta+T)\omega_{0}+\phi_{1}-\phi_{2}}
a_n^{\rm Coh}+C.C.\right],\label{eqPnRWA}\end{equation}\\
where
\begin{widetext}
$$p_n^{\rm Cla}=\sum_{\alpha=0}^{n}\left\{\langle g|\tilde U^{(n-\alpha)}_{(t_3,t_2)}|g\rangle
\langle g|\tilde U^{(\alpha)}_{(t_1,0)}|\tilde g\rangle +
e^{-\Delta\Gamma}\langle g|\tilde
U^{(n-\alpha)}_{(t_3,t_2)}|e\rangle \langle e|\tilde
U^{(\alpha)}_{(t_1,0)}|\tilde
g\rangle\right\}+\sum_{\alpha=0}^{n-1}(1-e^{-\Delta\Gamma})\langle
g|\tilde U^{(n-\alpha-1)}_{(t_3,t_2)}|g\rangle \langle e|\tilde
U^{(\alpha)}_{(t_1,0)}|\tilde g\rangle$$
\begin{equation}
 + \sum_{\alpha=0}^{n-1}\left\{\langle
e|\tilde U^{(n-\alpha-1)}_{(t_3,t_2)}|g\rangle \langle g|\tilde
U^{(\alpha)}_{(t_1,0)}|\tilde g\rangle +e^{-\Delta\Gamma}\langle
e|\tilde U^{(n-\alpha-1)}_{(t_3,t_2)}|e\rangle \langle e|\tilde
U^{(\alpha)}_{(t_1,0)}|\tilde
g\rangle\right\}+\sum_{\alpha=0}^{n-2}(1-e^{-\Delta\Gamma})\langle
e|\tilde U^{(n-\alpha-2)}_{(t_3,t_2)}|g\rangle \langle e|\tilde
U^{(\alpha)}_{(t_1,0)}|\tilde g\rangle\label{PnclaRWA}\end{equation}
and
\begin{equation}a_n^{\rm Coh}=\sum_{\alpha=0}^{n}\langle
g|\tilde U^{(n-\alpha)}_{(t_3,t_2)}|c\rangle \langle c|\tilde
U^{(\alpha)}_{(t_1,0)}|\tilde g\rangle  +
\sum_{\alpha=0}^{n-1}\langle e|\tilde
U^{(n-\alpha-1)}_{(t_3,t_2)}|c\rangle \langle c|\tilde
U^{(\alpha)}_{(t_1,0)}|\tilde
g\rangle,\label{AncohRWA}\end{equation}
with $|\tilde g\rangle = A_{(0,\phi_1,\omega_L)}| g\rangle$.
\section{Appendix C}

   Here we show the derivation of exact expressions for $P_1$ and $P_2$ for two equal square
pulses obtained within RWA .\\

 \textbf{For $P_1$ :}
 The one-photon-propagator in (0,t) may be decomposed to
the sum of 4 different terms
\begin{equation}
U^{(1)}_{(t,0)}=\sum_{k=0}^1 \left[
U^{(\alpha)}_{(t,t_3)}U^{(1-\alpha)}_{(t_3,t_2)}U^{(0)}_{(t_2,t_1)}U^{(0)}_{(t_1,0)}
+
U^{(0)}_{(t,t_3)}U^{(0)}_{(t_3,t_2)}U^{(\alpha)}_{(t_2,t_1)}U^{(1-\alpha)}_{(t_1,0)}
\right] \label{eqU1exact}
\end{equation}
Inserting the closure relation Eq.~(\ref{closure}) between every two
propagators of Eq.~(\ref{eqU1exact}) and applying RWA we obtain the
probability of emitting a single photon using the trajectories
notation
$$p^{\rm Cla}_1 = \sum_{\alpha=0} ^1 \left\{ \langle g |\tilde U^{(\alpha)}_{t_3,t_2} | g \rangle \langle g |\tilde U^{(1-\alpha)}_{t_1,0} | g \rangle
+   \langle g |\tilde U^{(\alpha)}_{t_3,t_2}| e \rangle \langle e
|\tilde U^{(1-\alpha)}_{t_1,0} | g \rangle e^{- \Gamma \Delta }
\right\}
$$
\begin{equation}+\langle g |\tilde U^{(0)}_{t_3,t_2} | g \rangle\langle e |\tilde U^{(0)}_{t_1,0} | g \rangle \left( 1 - e^{ - \Gamma \Delta  } \right)
+\langle e |\tilde U^{(0)}_{t_3,t_2} | e \rangle \langle e |\tilde
U^{(1-\alpha)}_{t_1,0}| g \rangle e^{ - \Gamma \Delta} + \langle e
|\tilde U^{(0)}_{t_3,t_2}| g\rangle \langle g |\tilde
U^{(0)}_{t_1,0} | g \rangle. \label{eqponeCla}
\end{equation}
\begin{equation}
a^{\rm Coh}_1= \sum_{\alpha=0} ^1 \langle g|\tilde
U^{(\alpha)}_{t_3,t_2} | c \rangle \langle c |\tilde
U^{(1-\alpha)}_{t_1,0} | g \rangle + \langle e |\tilde
U^{(0)}_{t_3,t_2} |c\rangle \langle c | \tilde U^{(0)}_{t_1,0} | g
\rangle. \label{eqponeCoh}
\end{equation}
After some tedious algebra using Eqs.~(\ref{eqLoOBE}),
(\ref{eqLRWA}), (\ref{eqtildUn}), (\ref{eqtildeGreenf}) we finally
obtain :
\begin{equation}
P_1 = a_1+b_1 e^{-\Delta}+ c_1
   e^{-{\Delta \over 2}} \cos \left [\omega_0 \left (T + \Delta \right ) \right ].
   \label{P1}
\end{equation}
Where
\begin{equation}
   a_1 = \frac{e^{-T}}{16 y^7} \left(1-y^2\right) \left[a_{11}+a_{12} \cosh \left(\frac{T
   y}{2}\right)+a_{13}\sinh \left(\frac{T y}{2}\right) +a_{14}\cosh (T y) +  a_{15}\sinh (T y)\right].
\end{equation}
\begin{equation}y = \sqrt{1-4 \Omega^2},\end{equation}
\begin{eqnarray*}
&&a_{11}= y\left(y^2-1\right) \left[4 \left(y^2-3\right)+3 T\left(y^2-1\right)\right], a_{12}= -2 y \left(T y^4+8y^2-T-16\right),a_{13}= 4 \left[-(T+3) y^4+(T+4) y^2+3 \right], \\
&&a_{14}= (T+4) y^5+6 T y^3+(T-20) y, a_{15}= 2 \left[(2 T+5)y^4+2 (T-5) y^2-3\right]. \\
\end{eqnarray*}

\begin{equation}
b_1 = \frac{e^{-T}}{16 y^7}\left(1-y^2\right)^3 \left[-3 T y+2 T y
\cosh \left(\frac{T
   y}{2}\right) +12 \sinh \left(\frac{Ty}{2}\right) +T y \cosh (T y) -6 \sinh (T y)\right].
\end{equation}
And
\begin{equation}
c_1 = \frac{e^{-T}}{16 y^7} \left(1-y^2\right) \left[c_{11}+
   c_{12}\cosh \left(\frac{T y}{2}\right)+ c_{13}\sinh \left(\frac{T
   y}{2}\right) + c_{14}\cosh (T y)+  c_{15}\sinh(T y)\right].
\end{equation}
\begin{eqnarray*}
&&c_{11}= -2 y \left[2\left(y^4-3\right)+T \left(y^4+2y^2-3\right)\right],c_{12}= 4 y \left[(T+4) y^2-T-8\right],c_{13}=4 \left[Ty^4+(2-T) y^2-6\right], \\
&&c_{14}=2 y \left((T+2)y^4-8 y^2-T+10\right),c_{15}= 4 \left[T y^4 -(T+1) y^2 + 3\right]. \\
\end{eqnarray*}

\textbf{For $P_2$ :}\\

  The two-photon-propagator may be decomposed to the sum of 10 terms.
\begin{equation}
U^{(2)}_{(t,0)}=\sum_{\alpha=0}^2 \left[
U^{(0)}_{(t,t_3)}U^{(\alpha)}_{(t_3,t_2)}U^{(0)}_{(t_2,t_1)}U^{(2-\alpha)}_{(t_1,0)}+
U^{(\alpha)}_{(t,t_3)}U^{(0)}_{(t_3,t_2)}U^{(2-\alpha)}_{(t_2,t_1)}U^{(0)}{(t_1,0)}\right]
+  \sum_{\beta,\alpha=0}^1
U^{(\beta)}_{(t,t_3)}U^{(\alpha)}_{(t_3,t_2)}U^{(1-\beta)}_{(t_2,t_1)}U^{(1-\alpha)}_{(t_1,0)}
\label{eq2exact}
\end{equation}
( Since $U^{(2)}_{t_2,t_1}=0$ there are only 8 non-zero terms.)
 This leads to the following expressions for $P^{\rm Cla}_2$ and $A^{\rm Coh}_2$ :
$$p_2^{\rm Cla}= \sum_{k=0} ^2 \left\{ \langle g |\tilde  U^{(k)}_{(t_3,t_2)} | g \rangle \langle g |\tilde  U^{(2 - k)} _{(t_1,0)}  | g \rangle + \langle g |\tilde U^{(k)}_{(t_3,t_2)} | e \rangle \langle e |\tilde  U^{(2 - k)} _{(t_1,0)}  | g \rangle e^{- \Gamma \Delta }
 \right\}+ \langle e | \tilde U^{(0)}_{(t_3,t_2)} | g \rangle\langle e |\tilde U^{(0)} _{(t_1,0)} | g \rangle \left( 1 - e^{ - \Gamma \Delta  } \right)$$\\
$$ + \sum_{k=0} ^1 \left\{\langle g |\tilde U^{(k)}_{(t_3,t_2)}| g \rangle \langle e |\tilde U^{(1 - k)} _{(t_1,0)} | g \rangle  \left( 1 - e^{ - \Gamma \Delta  } \right)+ \langle e | \tilde U^{(k)}_{(t_3,t_2)}| g \rangle \langle g |\tilde U^{(1 - k)} _{(t_1,0)}  | g \rangle + \langle e |\tilde U^{(k)}_{(t_3,t_2)}| e \rangle \langle e | \tilde U^{(1 - k) }_{(t_1,0)}  | g \rangle e^{- \Gamma \Delta } \right\}  $$
\begin{equation}
a_2^{\rm Coh}=\sum_{k=0} ^2 \langle g|\tilde U^{(k)}_{(t_3,t_2)}| c
\rangle \langle c |\tilde  U^{(2 - k)} _{(t_1,0)} | g \rangle +
 \sum_{k=0} ^1 \langle e|\tilde U^{(k)}_{(t_3,t_2)} | c \rangle \langle c |\tilde U^{(1 - k)} _{(t_1,0)}  | g \rangle  + \mbox{C.C.}
\label{eqpone}
\end{equation}
Calculating the matrix elements with the help of Mathematica yields:
\begin{equation}
P_2 = a_2+ b_2 e^{-\Delta}+ c_2e^{-{\Delta \over 2}} \cos
\left[\omega_0 \left (T + \Delta \right ) \right]. \label{P2}
\end{equation}
where
\begin{equation}
a_2 = \frac{e^{-T}}{64 y^{10}}
\left(y^2-1\right)^2\left[a_{21}+a_{22} \cosh
\left(\frac{Ty}{2}\right)+ a_{23} \sinh\left(\frac{T
y}{2}\right)+a_{24} \cosh (T y)+ a_{25} \sinh (Ty)\right].
\end{equation}
%
%
%
\begin{eqnarray*}
&&a_{21} = \left(9 T^2+40 T+24\right) y^6-2 \left(9 T^2+56T+17\right) y^4+9 \left(T^2+8 T-12\right) y^2+126,  \\
&&a_{22} = \left(T^2-32\right) y^6+4 (T+32) y^4-\left(T^2+20T+64\right) y^2-192, \\
&&a_{23} = 2 y \left[\left(y^2-1\right) T^2 y^2+T \left(3 y^4-8y^2-3\right)-4 \left(8 y^4-39 y^2+51\right)\right], \\
&&a_{24} = \left(T^2+8 T+8\right) y^6+2 \left(3 T^2-6 T-47\right)y^4+\left(T^2-52 T+172\right) y^2+66,\\
&&a_{25} = y \left[4 \left(y^2+1\right) T^2 y^2+\left(17 y^4-58y^2-15\right) T-4 \left(4 y^4+9 y^2-51\right)\right]. \\
\end{eqnarray*}

\begin{equation}
b_2 = \frac{e^{-T}}{64 y^{10}} \left(y^2-1\right)^2 \left[b_{21}+
b_{22} \cosh \left(\frac{Ty}{2}\right)+ b_{23} \sinh\left(\frac{T
y}{2}\right)+ b_{24} \cosh (T y)+b_{25} \sinh (Ty)\right].
\end{equation}
\begin{eqnarray*}
&&b_{21} = 3\left(3 T^2-8\right) y^6-18 \left(T^2-7\right) y^4+9\left(T^2-28\right) y^2+126,b_{22} = -\left(T^2-32\right) y^6+2 \left(T^2-96\right)y^4-\left(T^2-384\right) y^2-192, \\
&&b_{23} = -6 T y \left(y^2-1\right)^2,b_{24} = \left(T^2-8\right) y^6-2 \left(T^2-33\right)y^4+\left(T^2-132\right) y^2+66,b_{25} = -15 T y \left(y^2-1\right)^2. \\
\end{eqnarray*}
And
\begin{equation}
c_2 = \frac{e^{-T}}{32 y^{10}} \left(y^2-1\right)^2\left[c_{21}+
c_{22} \cosh \left(\frac{Ty}{2}\right)+ c_{23} \sinh\left(\frac{T
y}{2}\right)+ c_{24} \cosh (T y)+ c_{25} \sinh (Ty)\right],
\end{equation}
\begin{eqnarray*}
&&c_{21} = -T (T+4) y^6+2 \left(5 T^2+12 T-15\right) y^4-3 \left(3T^2+12 T-44\right) y^2-126, \\
&&c_{22} = -\left(T^2+2 T-32\right) y^4+\left(T^2+10 T-160\right)y^2+192, \\
&&c_{23} = -\left(T^2-32\right) y^5+\left(T^2+2 T-156\right) y^3+6(T+34) y,\\
&&c_{24} = T (T+4) y^6-2 (11 T+1) y^4+\left(-T^2+26 T+28\right)y^2-66, \\
&&c_{25} = y \left[2 \left(y^2-1\right)T^2  y^2+T \left(-3 y^4-4y^2+15\right)-2 \left(8 y^4-39 y^2+51\right)\right]. \\
\end{eqnarray*}
\end{widetext}


\begin{thebibliography}{99}

\bibitem{MukamelB} S. Mukamel  {\em Principles of Nonlinear Optical
Spectroscopy} Oxford Univ. Press. Oxford (1995).

\bibitem{vanDijk} Erik M.H.P van Dijk et al  {\em Phys. Rev. Lett.}
{\bf 94} 078302 (2005).

\bibitem{SB} F. Shikerman, E. Barkai {\em
 http://arxiv.org/abs/0705.4028} (2007).

\bibitem{Santori}C. Santori, D. Fattal, J. Vu$\hat{c}$kovi$\acute{c}$, G. S. Solomon  and Y. Yamamoto
 {\em Nature} {\bf 419}, 594-597 (2002).

\bibitem{Hong} C.K. Hong, Z. Y. Ou, L. Mandel  {\em Phys. Rev. Lett.}
{\bf 59} 2044-2046 (1987).

\bibitem{Knill} E. Knill, R. Laflamme and G. J. Milburn  {\em Nature}
{\bf 409} 46-52 (2001).

\bibitem{Shih} Y.H. Shih, C. O. Alley  {\em Phys. Rev. Lett.}
{\bf 61} 2921-2024 (1988).

\bibitem{Katz} N. Katz, M. Ansmann, R. C. Bialczak, E. Lucero,
R. McDermott, M. Neeley, M. Steffen, E. M. Weig, A. N. Korotkov {\em
Sience} {\bf 312}, 1498-1500 (2006).

\bibitem{Bouwmeester} D. Bouwmeester, A. Ekert, A. Zelinger {\em the Physics of Quantum Information}49-92 Springer, Berlin (2000).

\bibitem{Zoller}  P. Zoller, M. Marte, and D. F. Walls  {\em Phys. Rev. A} {\bf 35} 198 (1987).

\bibitem{Mollow} B. R. Mollow,  {\em Phys. Rev. A}  {\bf 12}, 1919
(1975).

\bibitem{Orrit} C. Brunel, B. Lounis, P. Tamarat and M. Orrit
{\em Phys. Rev. Lett.} {\bf 83} 2722 (1999).

\bibitem{BarkaiPRL}  E. Barkai, Y.J. Jung, and R. Silbey,
{\em Phys. Rev. Lett.} {\bf 87}, 207403 (2001).

\bibitem{BarkaiRev} E. Barkai, Y. Jung and  R. Silbey
 {\em Annual Review of Physical Chemistry} {\bf 55}, 457 (2004).

\bibitem{Mukamel} S. Mukamel
 {\em Phys. Rev. A} {\bf 68} 063821 (2003).

\bibitem{Cao} J.S. Cao
{\em  J. of  Phys. Chem. B} {\bf  110}  19040 (2006)

\bibitem{Xie} H. Yang, X.C. Xie  {\em J. of Chem. Phys.} {\bf117} 10965
( 2002).

\bibitem{Goppich} I. Gopich, A. Szabo
{\em  J. of Chemical Physics} {\bf  122} 014707 (2005).

\bibitem{Brown} Y. Zheng, F. L. H. Brown  {\em Phys. Rev. Lett.}  {\bf 90}
238305  (2003).

\bibitem{Plenio} M. B. Plenio and P. L. Knight,  {\em Rev. of Mod. Phys.}
{\bf70} 101-144 (1998).

\bibitem{CT} C. C. Tannoudji, J. Dupont-Roc, G. Grynberg  {\em Atom-Photon
Interactions}, John Wiley (1992).


\bibitem{Rozkov} I. Rozhkov, E. Barkai  {\em J. of Chem. Phys.}
{\bf 123} 074703 (2005).

\bibitem{remark} Reperesenting $\sigma^{(n)}$ as a $2\times 2$ matrix,
Eq.~(\ref{eqPn}) is just the trace of the matrix. While some authors
prefer the presentation of $\sigma^{(n)}$ as a matrix,  we follow
Zheng and Brown \cite{Brown} presentation of $\sigma^{(n)}$ as a
vector.


\bibitem{Yong} Y. He, E. Barkai
{\em Physical Chemistry Chemical Physics} {\bf8}, 5056 (2006).

\bibitem{Yong1} Y. He, E. Barkai  {\em Phys. Rev. A} {\bf74}, 011803(R)
(2006).

\end{thebibliography}
\end{document}